\documentclass[article]{IEEEtran}
\IEEEoverridecommandlockouts
\usepackage{nicematrix}
\usepackage{arydshln} 
\usepackage{mathtools}
\usepackage[dvipsnames]{xcolor}
\usepackage{cuted}

\DeclareFontFamily{U}{boondoxuprscr}{\skewchar \font =45}
\DeclareFontShape{U}{boondoxuprscr}{m}{n}{
    <-> BOONDOXUprScr-Regular}{}
\DeclareFontShape{U}{boondoxuprscr}{b}{n}{
    <-> BOONDOXUprScr-Bold}{}

\newcommand\kay{\text{\fontencoding{U}\fontfamily{boondoxuprscr}\fontshape{n}\selectfont k}}

\usepackage{booktabs}
\usepackage{array}
\usepackage{tabularray}
\usepackage{arydshln}
\usepackage{arydshln,leftidx,mathtools}

\usepackage{siunitx}
\usepackage{amsmath,amssymb,amsfonts}
\usepackage{algorithmic}
\usepackage{graphicx}
\usepackage{textcomp}
\usepackage{xcolor}
\usepackage{amsmath}
\usepackage{amssymb}
\usepackage{mathtools}
\usepackage{amsthm}
\usepackage{amsfonts}
\usepackage{stackengine}

\usepackage{float}
\usepackage{amsmath,amsthm,amsfonts,amscd,amssymb, arydshln}
\usepackage[mathscr]{eucal}
\usepackage{graphics,graphicx,multicol}
\usepackage{epsfig}
\usepackage{diagbox}
\usepackage{subfigure}
\usepackage{stfloats}
\usepackage{latexsym}
\usepackage{amsfonts}
\usepackage{cite}
\usepackage{algorithm,algorithmic}
\usepackage{color}
\usepackage{xcolor}
\usepackage{multirow}
\usepackage{booktabs}

\usepackage[utf8]{inputenc} 
\usepackage[T1]{fontenc}

\DeclareMathOperator*{\argmin}{arg\,min}
\title{}
\author{}
\date{}

\def\BibTeX{{\rm B\kern-.05em{\sc i\kern-.025em b}\kern-.08em
    T\kern-.1667em\lower.7ex\hbox{E}\kern-.125emX}}
\begin{document}

\title{Pilot-Free Predictive Multi-User Beamforming via Sensing Management in Cell-Free Networks
\thanks{This work was supported by the SUCCESS project funded by the Swedish Foundation for Strategic Research. A preliminary version of this work was presented at the IEEE International Symposium on Personal, Indoor and Mobile Radio Communications,
1-4 September 2025.}
}
\author{Eren Berk Kama,~\IEEEmembership{Student Member,~IEEE}, Murat Babek Salman,~\IEEEmembership{Member,~IEEE},\\ Isaac Skog,~\IEEEmembership{Member,~IEEE}, and Emil Bj{\"o}rnson,~\IEEEmembership{Fellow,~IEEE}
\thanks{Eren Berk Kama, Isaac Skog and Emil Bj{\"o}rnson are with the Division of Communication Systems, KTH Royal Institute of Technology, Stockholm, Sweden (e-mail: ebkama@kth.se; skog@kth.se; emilbjo@kth.se). {Murat Babek Salman is with Ericsson (e-mail: murat.salman@ericsson.com).}
}}

\maketitle
\begin{abstract}
This paper presents a sensing management framework for integrated sensing and communications (ISAC) within cell-free massive multiple-input multiple-output (MIMO) systems to reduce pilot-based channel state information (CSI) acquisition overhead. Conventional communication systems rely on frequent channel estimation procedures that impose significant signaling overhead, consuming valuable time-frequency resources. To address this inefficiency, we propose a state-based architecture that partitions users into communication and sensing groups based on service requirements. When users are not requesting data, the system utilizes sensing capabilities to track their location. Upon receiving a communication request, the system transitions to communication mode, leveraging the tracked state for predictive beamforming to eliminate the need for uplink pilot training. We develop an extended Kalman filter (EKF) based tracking algorithm coupled with adaptive resource allocation strategies. Furthermore, we analyze the impact of inter-target interference and design a sensing management protocol that performs sensing operations only when necessary to maintain the accuracy of user location estimates. Simulation results demonstrate that the proposed EKF-based tracking and sensing management can support predictive beamforming with downlink spectral efficiency close to the perfect-CSI case, while requiring sensing only occasionally after an initial convergence period. The results also indicate that this performance is robust in a cell-free massive MIMO setup and can be achieved with practical sensing waveforms.
\end{abstract}


\section{Introduction}

In recent years, the combination of wireless communications and sensing has attracted considerable attention \cite{10217169}. In the current infrastructure, radar and communication systems are operated independently, each with its dedicated hardware, deployment, and spectrum \cite{9737357}. However, integrated sensing and communications (ISAC) seeks to unify these functionalities to achieve multifunctional wireless systems with joint resource allocation and enhanced performance for both services.

Pilot-based channel estimation for beamforming imposes a notable resource overhead. Prior work has addressed this issue by employing user tracking and predictive beamforming. \textcolor{black}{Bayesian and extended Kalman filter (EKF)-based tracking methods have been applied to predict target motion from radar echoes, enabling predictive beamforming in vehicular networks \cite{9246715}, massive multi-input multi-output (MIMO) vehicle-to-infrastructure communication \cite{9171304}, three-dimensional drone tracking \cite{10214383}, and wireless sensor network localization \cite{khan2014localization}. Further studies have investigated sensing-assisted beam tracking for extended targets with dynamic beamwidth adaptation \cite{du2022sensing}, incorporated non-line-of-sight (NLoS) identification to improve robustness \cite{10278781}, and exploited multipath echoes to enhance prediction accuracy \cite{10872824}.} Deep learning techniques have also been proposed to predict beamforming matrices, enhancing communication rates and reducing channel estimation overhead \cite{9791349,9492131}. Predictive beamforming strategies for distributed MIMO systems are presented in \cite{akçalı2025predictivebeamformingdistributedmimo}.

\textcolor{black}{Effective ISAC also relies on waveform design. Multicarrier phase coded waveforms have been shown to jointly support radar sensing and data transmission \cite{5776640,9354629}, with subsequent studies applying information- and estimation-theoretic criteria to adaptive subcarrier power allocation \cite{7970102}, limited-feedforward designs that account for control-signaling overhead in dual-functional radar--communications \cite{9420261}, and resource-occupancy-aware schemes that suppress ambiguity-function sidelobes and minimize delay--Doppler estimation errors while meeting communication constraints \cite{10548861}.}

\textcolor{black}{Recent works have explored cell-free massive MIMO ISAC systems in which coordinated access points (APs) perform both communication and sensing, including beamforming and power allocation designs \cite{10742291}, AP operation-mode selection and power control to satisfy communication quality-of-service and sensing requirements in mmWave cell-free ISAC \cite{yan2024communicate} and cell-free massive MIMO-assisted multi-target detection \cite{10901970}, comparisons of distributed and centralized multi-static sensing architectures \cite{10681604}, and graph neural network and deep reinforcement learning approaches for hybrid beamforming and beam selection \cite{du2025graph,zhang2025efficient}.}

Existing solutions typically assume simultaneous sensing of targets distinct from communication users, or track active users whose uplink pilots already provide inherent localization. \textcolor{black}{Motivated by bursty traffic patterns, this work considers a state-based ISAC architecture in cell-free massive MIMO, where users alternate between a sensing state during idle periods and a communication state upon data request, thereby transmitting data in bursts and eliminating the need for simultaneous radar and communication signals.} By leveraging position information obtained through user tracking during dedicated sensing periods, the proposed ISAC framework removes the need for channel estimation, thereby significantly reducing the associated overhead when users request access. \textcolor{black}{Moreover, the tracking filter's ability to predict user positions and quantify prediction uncertainty enables a sensing management method that activates sensing only when needed, thereby reducing the resources allocated to sensing.}

\begin{figure}[t!]
\centering
\includegraphics[scale=.1,width=8.7cm]{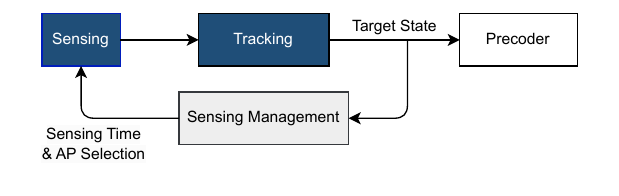}\vspace{-3mm}
\caption{Conceptual diagram of the proposed predictive beamforming method. Sensing and tracking provide an estimate of the user position information when needed, and the precoder is formed on demand. }
\label{fig:systemmodel}\vspace{-3mm}
\end{figure}

Fig.~\ref{fig:systemmodel} illustrates the considered system. \textcolor{black}{Within a cell-free massive MIMO architecture, we derive EKF state equations to track each user's position, velocity, and angle, enabling accurate angle prediction for precoding.} 
\textcolor{black}{To capture the coexistence of strong LoS paths and scattered NLoS components, we adopt a Rician fading channel model throughout the analysis.}

The main contributions of this paper are summarized as follows:
\begin{itemize}
\item We establish a system model for a multi-AP, multi-user, and multi-target ISAC network. Within this framework, we partition users based on their service requirements into communication and sensing groups, facilitating the simultaneous operation of multi-target tracking and downlink data transmission.
 \item We introduce a novel resource allocation framework that partitions subcarriers between sensing and communication users. Moreover, we examine two transmission strategies: the shared-waveform case, where a single waveform jointly supports sensing and communication, and the separate-waveform case, where distinct waveforms are allocated to each function.
\item We propose a predictive beamforming and power allocation algorithm designed for users in motion. The beamformer leverages target state information, obtained via tracking, to minimize the mean-squared error (MSE) of the predicted channel, thereby proactively managing interference. 
 \item We design a sensing management protocol that reduces signaling overhead by integrating target tracking with AP and user sensing time selection. This method allocates Rx and Tx APs, and the times that sensing needs to be done for each user, thereby replacing high-overhead channel estimation.
 \item We analyze the impact of inter-target interference on sensing performance and quantify how multi-target sensing affects communication spectral efficiency. Moreover, we characterize how the partitioning of spectral and power resources between sensing and communication affects the resulting performance.
\end{itemize}

This article significantly broadens the contribution of our preliminary work \cite{11275289}.
In particular, the journal version generalizes the system model to a multi-AP, multi-user, and multi-target setting and introduces an explicit service-based partitioning of users into sensing and communication groups. It further incorporates a resource allocation, where resources are partitioned between sensing, and communication users, and is further studied in terms of common dual-functional waveforms and separate sensing waveforms. To reflect propagation characteristics in distributed AP deployments, the analysis adopts a Rician fading model, and we derive downlink SE expressions that quantify the trade-offs introduced by the proposed methods. 

\begin{figure}[t!]
\centering
        \includegraphics[scale=.01,width=8.7cm]{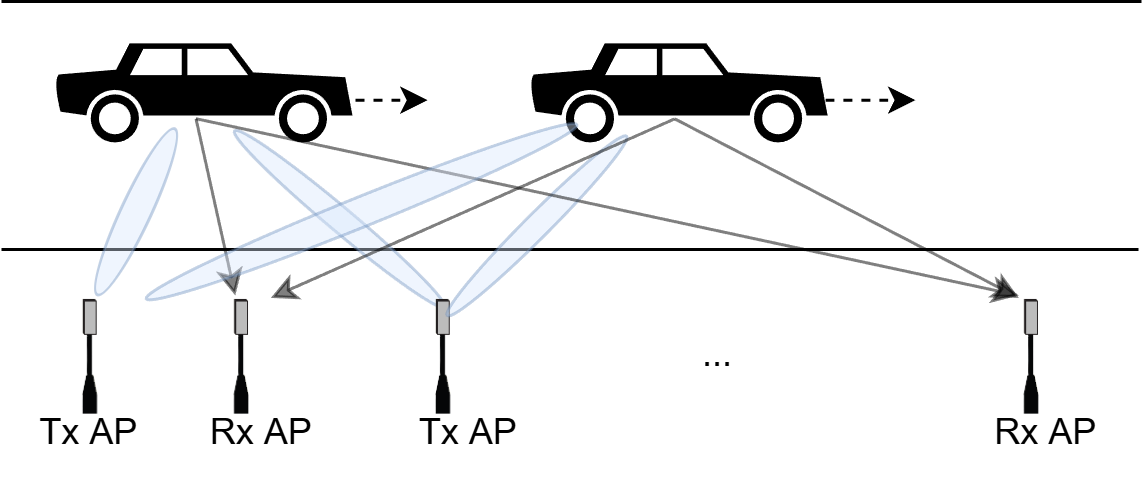}\vspace{-3mm}
    \caption{The considered system showing the set of chosen transmit and receive APs and the users.}
    \label{fig:APs and car system}\vspace{-3mm}
\end{figure}

\textit{Notation:} The superscripts $(\cdot)^{*}$,$(\cdot)^{\top}$, and $(\cdot)^{\mathrm{H}}$ denote the conjugate, transpose, and conjugate transpose, respectively.  Column vectors and matrices are denoted with boldface lowercase $\mathbf{x}$ and uppercase $\mathbf{X}$ letters. The expectation operation, variance, and covariance of a vector and matrix are denoted as $\mathbb{E}\{\cdot\}, \mathrm{var}(\cdot),\mathrm{cov}(\cdot)$. \textcolor{black}{The covariances are evaluated as $\mathbf{C}_{\mathbf{a}\mathbf{b}} 
= \mathbb{E}\!\left\{(\mathbf{a}-\mathbb{E}\{\mathbf{a}\})(\mathbf{b}-\mathbb{E}\{\mathbf{b}\})^{\mathrm H}\right\}
$ for $\mathbf{a}$ and $\mathbf{b}$ vectors.} The trace and determinant of a matrix are given by $\mathrm{tr}(\cdot)$ and $|\cdot|$, respectively. Convergence is shown with $\rightarrow$. The $L_{2}$ norm of a vector is given by $\Vert \mathbf{x} \Vert_{}$. The $N$-dimensional identity matrix is denoted by $\mathbf{I}_{N}$. $\mathbf{x} \sim \mathcal{N}_{\mathbb{C}}(\mathbf{0},\mathbf{R})$ denotes a circularly symmetric complex Gaussian random vector $\mathbf{x}$ with covariance matrix $\mathbf{R}$.
The vectorization of a matrix $\mathbf{X}$ is denoted by $\mathrm{vec}(\mathbf{X})$. The block diagonal operator is denoted by $\mathrm{blkdiag}[\cdot]$. 

\section{Communication and Sensing System Model}\label{sec:Communication and Sensing System Model}

We consider a cell-free massive MIMO ISAC system.
There are $L_{\mathrm{T}}$ APs, each with $N$ antenna elements, and $K$ single-antenna users. The APs are controlled by a central processing unit (CPU) and phase-synchronized to enable joint transmission and reception.

User traffic is inherently bursty in practice, leading to frequent transitions between active and idle states even when user applications are running continuously. A conventional network typically loses track of each user's location when the user switches to idle mode. To overcome this, we leverage sensing functionality so that when data is not transmitted, the APs use sensing signals to estimate and track the user position, enabling seamless service continuity. 
We consider the communication and tracking of multiple users simultaneously. 


We assume OFDM to be the common waveform for both communication and sensing operations. 
The OFDM signal at symbol time $b$ can be written as
\begin{equation} \label{eq:OFDM pulse}
\varsigma_{b}[m] =  \frac{1}{\sqrt{N_c}} \sum_{a=0}^{N_c-1} \gamma_{a,b} e^{j2\pi a \frac{m}{N_c}},
\end{equation}
where $N_c$ is the number of subcarriers  and $a$ and $b$ are the subcarrier and symbol indices, respectively. In a channel coherence block, the duration where the channel is assumed to be static, consisting of $N_{s}$ OFDM symbols, the time samples and symbols have the range $m=0,\ldots, N_{c}-1$ and $b=0,\ldots, N_{s}-1$, respectively. $\gamma_{a,b}$ are the transmitted symbols for the data signal or the code for sensing. To mitigate inter-symbol interference, the signal $\varsigma_{b}[m]$ is extended by appending a cyclic prefix (CP), which is formed by taking the last $N_{cp}$ samples of $\varsigma_{b}[m]$ and placing them at the beginning of the symbol creating an $N_{c}+N_{cp}$ length signal. The CP duration $N_{cp}$ is adjusted to be larger than the maximum round-trip delay of the targets considered in this paper. 

\begin{figure}[t]
\centering
        \includegraphics[scale=.5,width=8.7cm]{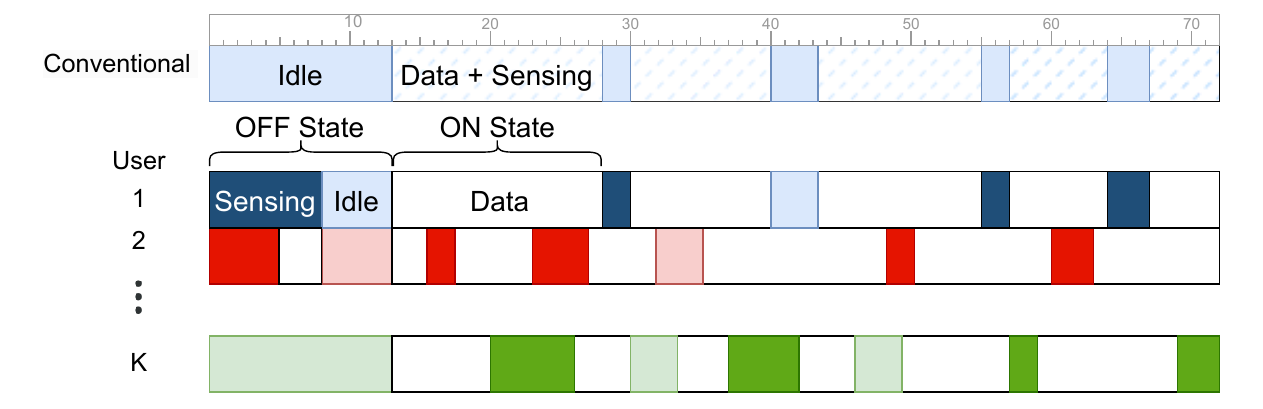}
    \caption{The conventional and proposed frame structures used in the transmissions from the CPU to the UE. Each user is denoted with a different color. Different-sized and placed blocks depict the separate communication and sensing operations for each user.}
    \vspace{-5.8mm}
    \label{fig:frame}
\end{figure}

We consider a state-based communication framework where data signals are transmitted to users upon request. The user remains in an \emph{OFF} state when it is not requesting data, during which no communication signals are transmitted to the user. Upon initiating a communication request, the user switches to the \emph{ON} state, triggering the transmission of communication signals. The state transition is controlled by the CPU. The communication signals are precoded towards the user based on the predicted user location, eliminating the need to allocate resources to estimate the channel at the AP side each time there is a communication request. 
In the conventional frame structure, considered in previous works such as \cite{akçalı2025predictivebeamformingdistributedmimo,9171304}, the system sends both data and sensing signals when there is a communication request from the user. Consequently, sensing is performed simultaneously with communication. By contrast, our proposed method utilizes the idle times in the conventional method for sensing.
The proposed and conventional frame structures are shown in Fig.~\ref{fig:frame} for comparison. The figure exemplifies the assumed state transitions in the proposed frame structure for all $K$ users, as well as the corresponding transitions between idle and communication states in the conventional structure. The time frame corresponds to that of the tracking filter, which is explained in Section ~\ref{sec: EKF}. We note that, in the proposed frame structure, sensing continues until a designated user tracking requirement is fulfilled, and it is repeated when the location information becomes outdated. The frame structure is constructed as follows: first, check whether sensing is needed; if not, check whether a communication request is present. This is performed simultaneously for all users. Then, Rx AP and communication and sensing power division operations are implemented. Furthermore, power allocation is performed for each user to enhance performance. We denote the set of users in the ON state as $\mathcal{K}_{C}$ and the set of users in the sensing state as $\mathcal{K}_{S}$.

\subsection{Communication Channel Model} \label{Sec: Communication Channel Model}
We assume that there is a Rician fading channel between AP $ l $ and user $k$, which is modeled as
\begin{align}
\mathbf{h}_{lk} =  \ \sqrt{\frac{K_{\mathrm{R}}}{1+K_{\mathrm{R}}}}\underbrace{e^{j\varphi_{lk}}\sqrt{\beta_{lk}} \, \mathbf{a}(\theta_{lk} )}_{=\bar{\mathbf{h}}_{lk}} + \sqrt{\frac{1}{1+K_{\mathrm{R}}}}\mathbf{h}_{\mathrm{NLOS},lk},
\end{align}
where $ \beta_{lk} $ is the large-scale fading coefficient, $e^{j\varphi_{lk}}$ with $\varphi_{lk} \sim \mathcal{U}[-\pi, \pi]$ is the phase-shift at the first antenna, $ \mathbf{a}(\theta_{lk} ) $ is the array response vector, and $\theta_{lk} $ is the angle of arrival in the azimuth plane at AP $ l $ from user $k$. Furthermore, $K_{\mathrm{R}}$ is the Rician-K factor that determines the strength of the line-of-sight (LOS) component compared to the non-line-of-sight (NLOS) component. The term $\beta_{lk}$ is the channel coefficient. Assuming that horizontal uniform linear arrays (ULA) are used by all APs, the array response vector between AP $l$ and user $k$ is $ \mathbf{a}(\theta_{lk} ) = [ 1,  e^{j \frac{2\pi}{\lambda} d \sin\theta_{lk}},  \ldots,  e^{j \frac{2\pi}{\lambda} (N-1)d \sin\theta_{lk}} ]^{\top}$,
where $ d $ is the antenna spacing and $\lambda$ the wavelength. For brevity, $\bar{\mathbf{h}}_{lk}$ denotes the LOS part of the channel. Moreover, the term $\mathbf{h}_{\mathrm{NLOS},lk}\sim \mathcal{N}_{\mathbb{C}}(\mathbf{0}, \mathbf{R}_{lk})$ is the NLOS part of the channel. The small-scale fading from NLOS  propagation has a covariance matrix $\mathbf{R}_{lk}$ that models the large-scale fading. The covariance matrix satisfies $\mathrm{tr}(\mathbf{R}_{lk})/N=\beta_{lk}$ within the Rician framework.


\subsection{Downlink Signal Transmission}\label{sec: Downlink Signal Transmission}
The downlink signal transmission is detailed in this section. We will separate it into two sub-schemes: \textit{shared subcarriers} (SS) and \textit{separate subcarriers} (SeS). We will further divide the shared subcarriers case based on whether the waveform is shared between communication signals and sensing signals. We name them as \textit{shared subcarriers with shared waveform} (SS-SW) and \textit{shared subcarriers with separate waveform} (SS-SeW). We will use a common notation for transmitted signals in this section. We explain the properties, differences, and inherent constraints of these methods in Section \ref{sec: Transmission Schemes}.

\subsubsection{Downlink Data Transmission}
Here, we will explain the downlink data transmission when the users initiate a communication request. All the APs in the transmitting APs' set $\mathcal{T}$ serve all the users through joint coherent transmission. 
If sensing is needed, communication occurs in the next epoch. The transmitted communication signal $\mathbf{x}_{\mathrm{C},\bar{l}}[m]\in \mathbb{C}^{N \times 1}$ from AP $\bar{l}$ at time instance $m$ can be written as
\begin{align}
    \mathbf{x}_{\mathrm{C},\bar{l}}[m]=  \sum_{i\in \mathcal{K}_{C}}\mathbf{w}_{\mathrm{C},\bar{l}i} \varsigma_{\mathrm{C},i}[m], 
\end{align}
where $\mathbf{w}_{\bar{l}i}\in \mathbb{C}^{N\times 1}$ is the precoder from AP $\bar{l}$ to the user $i$ and $\varsigma_{\mathrm{C},i}[m]$ is the data signal of the user. 

We assume that all the APs transmit the same stream to the user, thanks to coherent transmission. Furthermore, we assume that the data streams have unit power, i.e., $\mathbb{E}\{|\varsigma_{\mathrm{C},i}[m]|^2\}=1$. 
To achieve high SEs, as derived in Section \ref{sec: Transmission Schemes}, effective precoders must be designed. This, in turn, necessitates that the APs have accurate position information, which is acquired via sensing and tracking, as explained in later sections.

\subsubsection{Sensing Signal Transmission}\label{subsec: Sensing Transmitted}

When the user is in the OFF state, and there is a need to improve the location information through sensing, the system transitions to the sensing state. The same set of transmit APs $\mathcal{T}$ is used to transmit sensing signals. The transmitted sensing waveform from AP $\bar{l}$ at time instance $m$ can be written as
\begin{align}
    \mathbf{x}_{\mathrm{S},\bar{l}}[m]= \sum_{i\in\mathcal{K}_{S}}\mathbf{w}_{\mathrm{S},\bar{l}i} \varsigma_{\mathrm{S},\bar{l}i}[m], 
\end{align}
where $\mathbf{w}_{\mathrm{S},\bar{l}i}\in \mathbb{C}^{N \times 1}$ is the sensing precoder. Sensing precoders are also formed by using the tracked user location to improve the estimation performance. Here, $\varsigma_{\mathrm{S},\bar{l}i}[m]$ is the transmitted sensing waveform. We assume that the OFDM waveform, defined in \eqref{eq:OFDM pulse}, with a phase code is transmitted for sensing. We provide the details in Section \ref{sec: Transmission Schemes}.


\subsubsection{Transmit Power}
The total transmitted signal $\mathbf{x}_{\bar{l}}[m]=\mathbf{x}_{\mathrm{C},\bar{l}}[m]+\mathbf{x}_{\mathrm{S},\bar{l}}[m]$  should satisfy the transmit power constraint at AP $l$, which we define as
\begin{align}
    &\mathbb{E} \left\{\Vert\mathbf{x}_{\bar{l}}[m] \Vert^{2} \right\}=\nonumber\\  &\mathbb{E} \left\{\Big\Vert\sum_{i\in \mathcal{K}_{C}}\mathbf{w}_{\mathrm{C},\bar{l}i}\varsigma_{\mathrm{C},i}[m]+\sum_{i\in \mathcal{K}_{S}}\mathbf{w}_{\mathrm{S},\bar{l}i}\varsigma_{\mathrm{S},\bar{l}i}[m]\Big\Vert^{2}\right\} \leq \rho_{d},
\end{align}
where $\rho_d$ is the maximum AP transmit power.

\subsection{The Received Echo Signal }


Next, we describe the received echo signal for sensing. When both sensing and communication signals are transmitted to their respective users, the received signal $\mathbf{y}_{l,b}[m]\in \mathbb{C}^{N \times 1}$ at an Rx AP $l$, at symbol $b$ and time instant $m$ is given by
\begin{align}
&\mathbf{y}_{l,b}[m] = \sum_{k\in\mathcal{K}_{S}} \sum_{\bar{l}\in\mathcal{T}} \bar{\sigma}_{l\bar{l}k} \mathbf{h}_{lk}  \mathbf{h}_{\bar{l}k}^{\top}\Bigg[ \sum_{k'\in\mathcal{K}_{S}}\mathbf{w}_{\mathrm{S},\bar{l}k'}\breve{\varsigma}_{\mathrm{S},\bar{l}k',b}[m](\boldsymbol{\eta}_{l\bar{l}k})  \nonumber \\&
  \nonumber  \\&+\sum_{i\in\mathcal{K}_{C}}  \mathbf{w}_{\mathrm{C},\bar{l}i} \breve{\varsigma}_{\mathrm{C},\bar{l}{i},b}[m](\boldsymbol{\eta}_{l\bar{l}k}) \Bigg]+  \mathbf{n}_{l,b}[m],
\label{eq:single_snapshot_received_signal}
\end{align}
where the delayed and Doppler-shifted sensing waveform is 
\begin{align}
    &\breve{\varsigma}_{\mathrm{S},\bar{l}k,b}[m](\boldsymbol{\eta}_{l\bar{l}k})=\nonumber\\&e^{j2\pi  b T_{\mathrm{sym}} \nu_{l\bar{l}k}} \frac{1}{\sqrt{N_{\mathrm{c}}}} \sum_{a=0}^{N_{\mathrm{c}}-1} \gamma_{a,b}^{(\bar{l}k)} e^{j2\pi a \frac{m}{N_{\mathrm{c}}}} e^{-j2\pi a \Delta_{f} \tau_{l\bar{l}k}}
\end{align}
and the communication counterpart of it $\breve{\varsigma}_{\mathrm{C},\bar{l}i,b}[m](\boldsymbol{\eta}_{l\bar{l}k})$ is defined in the same way with data code sequences.

The first term in the brackets in \eqref{eq:single_snapshot_received_signal} is the radar return signal, and the second term is the reflections of communication signals from the sensing targets. The vector $\boldsymbol{\eta}_{l\bar{l}k} = [\tau_{l\bar{l}k}, \nu_{l\bar{l}k}]^{\top} \in \mathbb{R}^{2}$ stacks
$\tau_{l\bar{l}k}$ and $\nu_{l\bar{l}k}$ values, which are the propagation delay and the Doppler shift between the Tx AP $\bar{l}$ and the Rx AP $l$ via the target $k$, and $\Delta_{f}$ is the subcarrier separation. The code assigned to user $k$ and transmitted from the AP $\bar{l}$ is denoted by $\gamma_{a,b}^{(\bar{l}k)}$. 
Note that the data signals vary over users but not over the APs due to coherent transmission; in other words, $\gamma_{a,b}^{(i)}$ is different for each user, not APs.
The receiver noise is temporally and spatially white and denoted as $\mathbf{n}_{l,b}[m] \sim \mathcal{N}_{\mathbb{C}}(\mathbf{0},\sigma_n^2\mathbf{I}_{N}) \ $$ \in \mathbb{C}^{N \times 1}$. 
Finally, $\bar{\sigma}_{l\bar{l}k}=\sqrt{\rho_d} \sqrt{\frac{ \lambda^{2}}{(4\pi)^{3}}}\sigma_{l\bar{l}k}$ is the collection of the scalar terms, where 
$\rho_d$ is the transmit power and $\sigma_{l\bar{l}k}$ is the complex reflection coefficient containing the radar cross section (RCS) $\tilde{\sigma}_{l\bar{l}k}$ of the user $k$ seen by Rx AP $l$. 
The symbol duration is $T_{\mathrm{sym}}=\frac{1}{ \Delta_{f}}+\frac{N_{\mathrm{cp}}}{N_{c} \Delta_{f}}$.

If we expand the received signal using the LOS and NLOS components of both the downlink and the uplink channels, we obtain
\begin{align}
&\mathbf{y}_{l,b}[m]
= \sum_{k\in\mathcal{K}_S}\sum_{\bar{l}\in\mathcal{T}}\Bigg[\sum_{k'\in\mathcal{K}_{S}}\breve{\varsigma}_{\mathrm{S},\bar{l}k',b}[m](\boldsymbol{\eta}_{l\bar{l}k})\Big[
(\alpha_{\mathrm{LL},l\bar{l}kk'}+\nonumber\\ &\alpha_{\mathrm{LN},l\bar{l}kk'}) \mathbf{a}(\theta_{lk})
+ \big(\alpha_{\mathrm{NL},l\bar{l}kk'}+\alpha_{\mathrm{NN},l\bar{l}kk'}\big)
\mathbf{h}_{\mathrm{NLOS},lk}
\Big]+ \nonumber\\ &\sum_{i\in\mathcal{K}_C}
\breve{\varsigma}_{\mathrm{C},\bar{l}i,b}[m](\boldsymbol{\eta}_{l\bar{l}k})\Big[
\big(\alpha_{\mathrm{LL},l\bar{l}ki}+\alpha_{\mathrm{LN},l\bar{l}ki}\big) \mathbf{a}(\theta_{lk})
+ \nonumber\\&\big(\alpha_{\mathrm{NL},l\bar{l}ki}+\alpha_{\mathrm{NN},l\bar{l}ki}\big) \mathbf{h}_{\mathrm{NLOS},lk}
\Big]\Bigg]
+ \mathbf{n}_{l,b}[m].\label{eq:single_snapshot_received_signal_open}
\end{align}
In the remainder of this section, we will explain the underlying relations using the first term inside the brackets in \eqref{eq:single_snapshot_received_signal_open}, which corresponds to the sensing return. The same relations will apply to the second term trivially.

In \eqref{eq:single_snapshot_received_signal_open}, we observe that a combination of inner products of LOS and NLOS channels and the precoders appears. We group these inside the channel coefficient terms and name them based on whether they are formed by LOS or NLOS channels at the Rx or Tx side, respectively. We write the corresponding channel coefficients as
\begin{align}
        \alpha_{\mathrm{LL}, l\bar{l}kk'} &= \bar{\sigma}_{l\bar{l}k} \frac{K_{\mathrm{R}}}{1+K_{\mathrm{R}}}e^{j\varphi_{lk}}\sqrt{\beta_{lk}} \left(\bar{\mathbf{h}}_{\bar{l}k}^{\top} \mathbf{w}_{\mathrm{S},\bar{l}k'}\right) \nonumber \\        \alpha_{\mathrm{LN}, l\bar{l}kk'} &= \bar{\sigma}_{l\bar{l}k} \frac{\sqrt{K_{\mathrm{R}}}}{1+K_{\mathrm{R}}}e^{j\varphi_{lk}}\sqrt{\beta_{lk}}\left(\mathbf{h}_{\mathrm{NLOS},\bar{l}k}^{\top}\mathbf{w}_{\mathrm{S},\bar{l}k'}\right)   \nonumber\\
        \alpha_{\mathrm{NL}, l\bar{l}kk'} &= \bar{\sigma}_{l\bar{l}k}  \frac{\sqrt{K_{\mathrm{R}}}}{1+K_{\mathrm{R}}}\left(\bar{\mathbf{h}}_{\bar{l}k}^{\top} \mathbf{w}_{\mathrm{S},\bar{l}k'}\right)   \nonumber\\
        \alpha_{\mathrm{NN}, l\bar{l}kk'} &= \bar{\sigma}_{l\bar{l}k}  \frac{1}{1+K_{\mathrm{R}}}\left(\mathbf{h}_{\mathrm{NLOS},\bar{l}k}^{\top}\mathbf{w}_{\mathrm{S},\bar{l}k'}\right) \label{eq: alpha calculation}.
\end{align}
For the communication signals, we replace the precoder term $\mathbf{w}_{\mathrm{S},\bar{l}k'}$ with the $\mathbf{w}_{\mathrm{C},\bar{l}i}$ term. We denote the channel coefficient with $\alpha_{(\cdot), l\bar{l}ki}$ for each of the counterpart terms.

Normally, the delay $\tau_{l\bar{l}k}$ and Doppler shift $\nu_{l\bar{l}k}$ are path-dependent. An NLOS path is longer than an LOS path, resulting in a larger delay. Similarly, if scattering occurs off a moving object, the Doppler shift can also differ. However, for a stationary environment with scatterers close to the transmitter or receiver, the differences in delay and Doppler shift between the LOS and NLOS paths originating from the same target $k$ are insignificant, often falling within the same delay-Doppler cell. In our analysis, we make this simplifying assumption, which allows us to use a single echoed waveform term $\breve{\varsigma}_{\mathrm{S},\bar{l}k}(\boldsymbol{\eta}_{l\bar{l}k})$ for all paths related to target $k$. 
In \eqref{eq: alpha calculation}, $\alpha_{\mathrm{LL}, l\bar{l}kk}$ is the channel coefficient for the signal of interest, and the rest of the terms are for the unintended returns.

\section{Multi-Static Sensing and Sensing Performance}

In this section, we present the multi-static sensing framework used in the network and detail the joint estimation of parameters, along with the associated lower bound on the estimation-error covariance. The network transmits sensing signals for target localization during the sensing state, and the received signal, expressed in \eqref{eq:single_snapshot_received_signal_open}, is used to obtain estimates of the delay, Doppler shift, and angle parameters. 

To jointly estimate the parameters, we formulate a compact signal model by stacking the received signal over all subcarriers and pulses. 
Subsequently, we stack the $N$-dimensional spatial snapshots $\mathbf{y}_{l,b}[m]\in\mathbb{C}^{N\times 1}$ into a single vector $\mathbf{y}_l\in\mathbb{C}^{N_c N_s N\times 1}$ as
\begin{align}
\mathbf{y}_l \triangleq \big[ \mathbf{y}_{l,0}[0]^{\top},\ldots,\mathbf{y}_{l,N_s-1}[N_c\!-\!1]^{\top}\big]^{\top}. \label{eq:vectorized_y}
\end{align}
Similarly, we stack the sensing waveform with temporal changes corresponding to transmitter $\bar{l}$ and target $k$. We define the vector $\breve{\boldsymbol{\varsigma}}_{\mathrm{S},\bar{l}k}(\boldsymbol{\eta}_{l\bar{l}k})\in \mathbb{C}^{N_c N_s \times 1}$ formed by stacking the discrete time-frequency samples $\breve{\varsigma}_{\mathrm{S},\bar{l}k,b}[m](\tau_{l\bar{l}k},\nu_{l\bar{l}k})$ according to the same slow-time and fast-time index pair used in \eqref{eq:vectorized_y}. The same stacking is applied to the data signals, which we denote by $\breve{\boldsymbol{\varsigma}}_{\mathrm{C},\bar{l}i}$. With these definitions, the received signal vector admits the decomposition in \eqref{eq:y_vec_model}, at the top of the next page.
\begin{figure*}
\begin{align}
&\mathbf{y}_l
= \sum_{k\in\mathcal{K}_S}\sum_{\bar{l}\in\mathcal{T}}\Bigg[
\sum_{k'\in\mathcal{K}_{S}}\breve{\boldsymbol{\varsigma}}_{\mathrm{S},\bar{l}k'}(\boldsymbol{\eta}_{l\bar{l}k})\otimes\Big[\big(\alpha_{\mathrm{LL},l\bar{l}kk'}+\alpha_{\mathrm{LN},l\bar{l}kk'}\big) \mathbf{a}(\theta_{lk})
+ \big(\alpha_{\mathrm{NL},l\bar{l}kk'}+\alpha_{\mathrm{NN},l\bar{l}kk'}\big) 
\mathbf{h}_{\mathrm{NLOS},lk}
\Big]\nonumber\\&+\sum_{i\in\mathcal{K}_C}
\breve{\boldsymbol{\varsigma}}_{\mathrm{C},\bar{l}i}(\boldsymbol{\eta}_{l\bar{l}k})\otimes\Big[
\big(\alpha_{\mathrm{LL},l\bar{l}ki}+\alpha_{\mathrm{LN},l\bar{l}ki}\big) \mathbf{a}(\theta_{lk})
+ \big(\alpha_{\mathrm{NL},l\bar{l}ki}+\alpha_{\mathrm{NN},l\bar{l}ki}\big) \mathbf{h}_{\mathrm{NLOS},lk}
\Big]\Bigg]
+ \mathbf{n}_l,\label{eq:y_vec_model}
\end{align}
\end{figure*}
In \eqref{eq:y_vec_model}, $\mathbf{n}_l\sim\mathcal{N}_{\mathbb{C}}(\mathbf{0},\sigma_n^2 \mathbf{I}_{N_c N_s N})$ is the additive white Gaussian noise vector. The terms involving NLOS components, particularly the one with the coefficient $\alpha_{\mathrm{NN},l\bar{l}k}$, which arises from the product of random variables, make the received signal non-Gaussian distributed.

Our goal is to obtain the delay, Doppler shift, angle, and channel coefficient of the target. Position information is mainly available in the LOS component of the receive channel. Furthermore, note that for $k'\neq k$ the precoders suppress the directions of the other sensing users, yielding significantly lower channel coefficients. Therefore, in the received signal, only the terms corresponding to the intended LOS-to-LOS sensing return are treated as unknowns to be estimated, and the other terms are treated as random disturbances. This reduces the number of parameters to be estimated and enables us to pool information across all temporal and spatial samples when estimating the parameters of interest. To that end, let 
\begin{align}
    &\boldsymbol{\Phi}_{lk} = [\boldsymbol{\tau}_{lk}, \boldsymbol{\nu}_{lk}, \theta_{lk}, \mathrm{Re}\{\boldsymbol{\alpha}_{\mathrm{LL},lkk}\},\mathrm{Im}\{\boldsymbol{\alpha}_{\mathrm{LL},lkk}\}]
\end{align}
be the parameter vector related to user $k$, where the row vectors $\boldsymbol{\tau}_{lk}, \boldsymbol{\nu}_{lk},\mathrm{Re}\{\boldsymbol{\alpha}_{\mathrm{LL},lkk}\}$, and $\mathrm{Im}\{\boldsymbol{\alpha}_{\mathrm{LL},lkk}\}$ contain the contributions of the respective parameters from all $\vert\mathcal{T}\vert$ Tx APs. We will use the indexing $\{\kay_1, \ldots, \kay_{\vert\mathcal{K}_{S}\vert}\}$ to denote the elements in the set of sensing users.
The parameter vector for all users is then
\begin{align}\label{eq: Parameter_matrix}
    \boldsymbol{\Phi}_{l} = [\boldsymbol{\Phi}_{l\kay_{1}},\ldots,\boldsymbol{\Phi}_{l\kay_{\vert\mathcal{K}_{S}\vert}} ]^{\top}.
\end{align}
To estimate these parameters, we adopt a weighted non-linear least-squares (WNLS) estimator. We employ WNLS as it relies only on the first- and second-order statistics of the received signal.
The mean of the received signal, $\boldsymbol{\mu}_l(\Phi_l) = \sum_{k\in\mathcal{K}_S}\sum_{\bar{l}\in\mathcal{T}}
\alpha_{\mathrm{LL},l\bar{l}kk}\big(\breve{\boldsymbol{\varsigma}}_{\mathrm{S},\bar{l}k}(\boldsymbol{\eta}_{l\bar{l}k})\otimes \mathbf{a}(\theta_{lk})\big)$, contains the deterministic LOS components used for estimation. We note that the other terms have zero mean due to the uniformly distributed phase terms in the channel coefficients. Treating the parameters to be estimated as unknown deterministic variables, the covariance matrix of the received signal is given by
\begin{align}
&\mathbf{R}_{\mathbf{y}_l}(\Phi_l)
= \sum_{k\in\mathcal{K}_S}\sum_{\bar{l}\in\mathcal{T}}
\Bigg[\sum_{k'\in\mathcal{K}_{S},k'\neq k}\mathbf{R}_{t,\bar{l}k',k}^{(\mathrm{S})} \otimes 
\bigg[
\mathbb{E}\big\{|\alpha_{\mathrm{LL},l\bar{l}kk'}+\nonumber\\&\alpha_{\mathrm{LN},l\bar{l}kk'}|^2\big\}\mathbf{a}(\theta_{lk})\mathbf{a}^{\mathrm{H}}(\theta_{lk})+ 
\mathbb{E}\left\{|\alpha_{\mathrm{NL},l\bar{l}kk'}+\alpha_{\mathrm{NN},l\bar{l}kk'}|^2\right\}
\nonumber\\&
\mathbf{R}_{\mathrm{NLOS},lk}
\bigg]  + \sum_{i\in\mathcal{K}_C}
\mathbf{R}_{t,\bar{l}i,k}^{(\mathrm{C})} \otimes 
\bigg[
\mathbb{E}\left\{|\alpha_{\mathrm{LL},l\bar{l}ki}+\alpha_{\mathrm{LN},l\bar{l}ki}|^2\right\}
\nonumber\\&\mathbf{a}(\theta_{lk})\mathbf{a}^{\mathrm{H}}(\theta_{lk})
+ \mathbb{E}\left\{|\alpha_{\mathrm{NL},l\bar{l}ki}+\alpha_{\mathrm{NN},l\bar{l}ki}|^2\right\}
\mathbf{R}_{\mathrm{NLOS},lk}
\bigg]\nonumber\\&+ \mathbf{R}_{t,\bar{l}k,k}^{(\mathrm{S})} \otimes 
\mathbb{E}\left\{|\alpha_{\mathrm{LN},l\bar{l}kk}|^2\right\}
\mathbf{a}(\theta_{lk})\mathbf{a}^{\mathrm{H}}(\theta_{lk}) \Bigg]  + \sigma_n^2\,\mathbf{I}_{N_c N_s N} .
\label{eq:R_full}
\end{align}
We observe that while each summand in \eqref{eq:R_full} is space-time separable, i.e., a Kronecker product of a temporal and a spatial covariance matrix, the overall covariance matrix $\mathbf{R}_{\mathbf{y}_l}$ is a sum of such products and does not admit a single Kronecker factorization. The matrices $\mathbf{R}_{t,\bar{l}k',k}^{(\mathrm{S})}$ are Toeplitz matrices containing the temporal autocorrelations of the transmitted waveforms for the sensing signals
\begin{align}
\mathbf{R}_{t,\bar{l}k',k}^{(\mathrm{S})}
&\triangleq \breve{\boldsymbol{\varsigma}}_{\mathrm{S},\bar{l}k'}(\boldsymbol{\eta}_{l\bar{l}k})\breve{\boldsymbol{\varsigma}}_{\mathrm{S},\bar{l}k'}^{\mathrm{H}}(\boldsymbol{\eta}_{l\bar{l}k}) \in \mathbb{C}^{(N_c N_s)\times(N_c N_s)}. \label{eq:Rt_def_matrix}
\end{align}
A similar expression holds for the data signal as $\mathbf{R}_{t,\bar{l}i,k}^{(\mathrm{C})}
\triangleq \breve{\boldsymbol{\varsigma}}_{\mathrm{C},\bar{l}i}(\boldsymbol{\eta}_{l\bar{l}k})\breve{\boldsymbol{\varsigma}}_{\mathrm{C},\bar{l}i}^{\mathrm{H}}(\boldsymbol{\eta}_{l\bar{l}k})$.

We can write the WNLS estimate of $\boldsymbol{\Phi}_{l}$ using the relation \eqref{eq:y_vec_model} as
\begin{align}\label{eq:Estimate_of_parameters}
    \hat{\boldsymbol{\Phi}}_{l}^{\mathrm{WNLS}}= \argmin_{\boldsymbol{\Phi}_{l} \in \mathbb{R}^{(4\vert\mathcal{T}\vert+1)\vert\mathcal{K}_{S}\vert\times1}} \Vert \mathbf{y}_{l} - \boldsymbol{\mu}_{l}(\boldsymbol{\Phi}_l) \Vert^{2}_{\mathbf{W}_{l}},
\end{align}
where the weighted norm is defined as $\Vert \mathbf{y}_{l} - \boldsymbol{\mu}_{l}(\boldsymbol{\Phi}_l) \Vert^{2}_{\mathbf{W}_{l}}\triangleq (\mathbf{y}_{l} - \boldsymbol{\mu}_{l}(\boldsymbol{\Phi}_l))^{\mathrm{H}} \mathbf{W}_{l} (\mathbf{y}_{l} - \boldsymbol{\mu}_{l}(\boldsymbol{\Phi}_l))$, and $\mathbf{W}_l=\mathbf{R}_{\mathbf{y}_l}\!\big(\boldsymbol{\Phi}_l^{}\big)^{-1}$ is the weighting matrix. The WNLS estimator attains the following distribution for large $N_c$ and $N_s$ \cite{ljung1987system}
\begin{align}
\hat{\boldsymbol{\Phi}}_{l}^{\mathrm{WNLS}}\stackrel{\textrm{asymp.}}\sim \mathcal{N}_{}(\boldsymbol{\Phi}_{l}^{*},\boldsymbol{\mathcal{C}}_{\boldsymbol{\Phi}_{l}}),
\end{align}
where $\boldsymbol{\Phi}_{l}^{*}$ is the true parameter vector and $\boldsymbol{\mathcal{C}}_{\boldsymbol{\Phi}_{l}}$ is the asymptotic covariance matrix (ACM) for the parameters in $\boldsymbol{\Phi}_{l}$. 
Let $\boldsymbol{\mathcal{A}}_{\boldsymbol{\Phi}_{l}}=\boldsymbol{\mathcal{C}}_{\boldsymbol{\Phi}_{l}}^{-1} $. Then, the entries of the matrix $\boldsymbol{\mathcal{A}}_{\boldsymbol{\Phi}_{l}}$ are given by \cite{ljung1987system}
\begin{align}
    \big[\boldsymbol{\mathcal{A}}_{\boldsymbol{\Phi}_{l}}\big]_{i,j}
    &= 2\,\mathrm{Re}\!\left\{\frac{\partial \boldsymbol{\mu}_l^{\mathrm{H}}(\Phi_l)}{\partial \phi_i}\,
    \mathbf{R}_{\mathbf{y}_l}^{-1}(\Phi_l)\,
    \frac{\partial \boldsymbol{\mu}_l(\Phi_l)}{\partial \phi_j}\right\} .
    \label{eq:slepian_bangs}
\end{align}
We are mainly interested in an ACM for the delay-Doppler and angle parameters for the WNLS method. 

For the sake of notational clarity, we group the temporal parameters delay and Doppler shift and stack them in $\boldsymbol{\eta}_{l} = [\boldsymbol{\tau}_{l}, \boldsymbol{\nu}_{l}]^{\top} \in \mathbb{R}^{2\vert\mathcal{T}\vert\vert\mathcal{K}_{S}\vert\times1}$. We also group all angles in $\boldsymbol{\theta}_{l}\in \mathbb{R}^{1 \times \vert\mathcal{K}_{S}\vert}$, and all complex channel coefficient components in $\boldsymbol{\alpha}_l=[\mathrm{Re}\{\boldsymbol{\alpha}_{\mathrm{LL},l}\},\mathrm{Im}\{\boldsymbol{\alpha}_{\mathrm{LL},l}\}] \in \mathbb{R}^{2 \vert\mathcal{T}\vert\vert\mathcal{K}_{S}\vert\times1}$.
With these groupings, the entries of $\boldsymbol{\mathcal{A}}_{\boldsymbol{\Phi}_{l}}$ can be written as
\begin{align}
\boldsymbol{\mathcal{A}}_{\Phi_l}
=
\begin{bmatrix}
\boldsymbol{\mathcal{A}}_{\boldsymbol{\eta}_l\boldsymbol{\eta}_l} & \boldsymbol{\mathcal{A}}_{\boldsymbol{\eta}_l\boldsymbol{\theta}_{l}} & \boldsymbol{\mathcal{A}}_{\boldsymbol{\eta}_l\boldsymbol{\alpha}_l} \\
\boldsymbol{\mathcal{A}}_{\boldsymbol{\theta}_{l}\boldsymbol{\eta}_l} & \boldsymbol{\mathcal{A}}_{\boldsymbol{\theta}_{l}\boldsymbol{\theta}_{l}} & \boldsymbol{\mathcal{A}}_{\boldsymbol{\theta}_{l}\boldsymbol{\alpha}_l} \\
\boldsymbol{\mathcal{A}}_{\boldsymbol{\alpha}_l\boldsymbol{\eta}_l} & \boldsymbol{\mathcal{A}}_{\boldsymbol{\alpha}_l\boldsymbol{\theta}_{l}} & \boldsymbol{\mathcal{A}}_{\boldsymbol{\alpha}_l\boldsymbol{\alpha}_l}
\end{bmatrix}.
\end{align}
The ACM for $[\boldsymbol{\eta}_{l}^{\top}\ \boldsymbol{\theta}_{l}]^{\top}$ can be found by using the Schur complement of the nuisance block. It reads
\begin{align}
\boldsymbol{\mathcal{C}}_{\boldsymbol{\eta}_l\boldsymbol{\theta}_{l}}
=
\Big(
&\begin{bmatrix}
\boldsymbol{\mathcal{A}}_{\boldsymbol{\eta}_l\boldsymbol{\eta}_l} & \boldsymbol{\mathcal{A}}_{\boldsymbol{\eta}_l\boldsymbol{\theta}_{l}} \\
\boldsymbol{\mathcal{A}}_{\boldsymbol{\theta}_{l}\boldsymbol{\eta}_l} & \boldsymbol{\mathcal{A}}_{\boldsymbol{\theta}_{l}\boldsymbol{\theta}_{l}}
\end{bmatrix}
- \nonumber \\
&\begin{bmatrix}
\boldsymbol{\mathcal{A}}_{\boldsymbol{\eta}_l\boldsymbol{\alpha}_l} \\
\boldsymbol{\mathcal{A}}_{\boldsymbol{\theta}_{l}\boldsymbol{\alpha}_l}
\end{bmatrix}
\boldsymbol{\mathcal{A}}_{\boldsymbol{\alpha}_l\boldsymbol{\alpha}_l}^{-1}
\begin{bmatrix}
\boldsymbol{\mathcal{A}}_{\boldsymbol{\alpha}_l\boldsymbol{\eta}_l} & \boldsymbol{\mathcal{A}}_{\boldsymbol{\alpha}_l\boldsymbol{\theta}_{l}}
\end{bmatrix}
\Big)^{-1}.
\label{eq:CRB_no_psi}
\end{align}
The ACM characterizes the fundamental lower bound on the estimation error variance, providing a theoretical benchmark that be utilized in Section \ref{Sec: Tracking} in the tracking framework to quantify the achievable localization accuracy.

\subsection{A Special Case}
Next, we will investigate a special case in which the ACM becomes block-diagonal, significantly reducing computational complexity. This happens with a space–time separable covariance model, which is possible only when all temporal covariances are proportional to a common matrix and all spatial covariances are co-linear. A separable covariance is in the form of $\mathbf{R}_{\mathbf{y}_l}(\boldsymbol{\Phi}_l)
= \mathbf{R}_{t,l}\otimes \mathbf{R}_{s,l}$. This happens when there is a single user and a single Tx AP. With this assumption, we drop the indices of each parameter and variable.


With the approximate space-time separable model, the angle and delay-Doppler ACMs are decoupled, resulting in a block diagonal ACM relation as in \cite{dogandzic2001cramer}
 \begin{align}    \boldsymbol{\mathcal{C}}_{\boldsymbol{\eta}_{l}\theta_{l}}=\begin{bmatrix}
         \boldsymbol{\mathcal{C}}_{\boldsymbol{\eta}_{l}} & \mathbf{0} \\ \mathbf{0}&\boldsymbol{\mathcal{C}}_{\theta_{l}}
     \end{bmatrix}.
 \end{align}
While not considered in the numerical results in this paper, this simplified form provides analytical insight and can simplify ACM computations in scenarios such as \cite{11275289}.
\section{User Tracking}\label{Sec: Tracking}

In this section, we build upon the estimation procedure developed in the previous section to develop a tracker for each user. The WNLS parameter estimates in \eqref{eq:Estimate_of_parameters} are employed as measurements in an extended Kalman filter (EKF) that tracks and predicts each user's state, from which the user's angle of arrival is derived. We assume that track association is handled. Therefore, an EKF can be applied for each user separately, and we drop the user indices in this section.

\subsection{Extended Kalman Filter} \label{sec: EKF}
Consider the scenario illustrated in Fig. \ref{fig:APs and car system}. We define the state vector of the user as
\begin{align}
    \mathbf{x}_{\kappa} = [p_{x}[\kappa] \ v_{x}[\kappa]]^{\top},
\end{align} 
where $p_{x}[\kappa]$ and  $v_{x}[\kappa]$ are the horizontal position and velocity of the user and $\kappa$ is the time index for the filter. Each $\kappa$ corresponds to $N_{s}$ OFDM symbols, and each OFDM symbol contains $N_{c}$ time samples. 
Furthermore, as described in Section~\ref{subsec: Sensing Transmitted}, a subset of APs is designated as the Rx APs. We use the indexing $\{\ell_1, \ldots, \ell_{L_\kappa}\}$ to denote the set of selected Rx APs at epoch $\kappa$. We only use the measurements obtained from them and define the measurement vector by stacking, for each Rx AP $\ell \in \mathcal{R}$, the estimated bistatic delay and Doppler shifts for all Tx APs $\bar{\ell}\in\mathcal{T}$ followed by the angle of arrival at Rx AP $\ell$
\begin{align}
&\mathbf{z}_{\kappa} = \big[
\underbrace{\hat{\tau}_{\ell_1\bar{\ell}_1}[\kappa], \ldots,\hat{\tau}_{\ell_1\bar{\ell}_{|\mathcal{T}|}}[\kappa], \hat{\nu}_{\ell_1\bar{\ell}_1}[\kappa], \ldots,\hat{\nu}_{\ell_1\bar{\ell}_{|\mathcal{T}|}}[\kappa], \hat{\theta}_{\ell_1}[\kappa]}_{\text{Rx }\ell_1},\nonumber\\
 &\ldots, \nonumber\\ &\underbrace{\hat{\tau}_{\ell_{L_\kappa}\bar{\ell}_1}[\kappa], \ldots, \hat{\tau}_{\ell_{L_\kappa}\bar{\ell}_{|\mathcal{T}|}}[\kappa],\hat{\nu}_{\ell_{L_\kappa}\bar{\ell}_1}[\kappa], \ldots,\hat{\nu}_{\ell_{L_\kappa}\bar{\ell}_{|\mathcal{T}|}}[\kappa], \hat{\theta}_{\ell_{L_\kappa}}[\kappa]}_{\text{Rx }\ell_{L_\kappa}}
\big]^{\top}.
\end{align}
Here, $\hat{\tau}_{\ell\bar{\ell}}[\kappa]$ and $\hat{\nu}_{\ell\bar{\ell}}[\kappa]$ are the estimated bistatic delay and Doppler shift for the Tx–Rx pair $(\bar{\ell}\rightarrow \ell)$ and the considered user, and $\hat{\theta}_{\ell}[\kappa]$ is the angle of arrival at Rx AP $\ell$. We note that 
uplink pilots could be fused into the measurement vector. However, they are unnecessary for the present sensing-centric design as the WNLS procedure yields direct range, Doppler shift, and angle estimations, rendering pilot-based measurements redundant.

Assuming that the user motion can be modeled by a constant velocity model, the state-space model that describes the system behavior is given by 
\begin{align}\label{eq:state space} \mathbf{x}_{\kappa+1} &= \mathbf{F}\mathbf{x}_{\kappa} + \mathbf{w}_{\kappa}, \quad \mathbf{w}_{\kappa} \stackrel {i.i.d.}\sim \mathcal{N}(\mathbf{0},\mathbf{Q}), \\ \mathbf{z}_{\kappa} &= \mathbf{h}(\mathbf{x}_{\kappa}) + \mathbf{v}_{\kappa}, \quad \mathbf{v}_{\kappa} \stackrel {i.d.}\sim \mathcal{N}(\mathbf{0},\mathbf{R}_{\kappa}). \end{align}
We note that the measurements are not independent; they are only identically distributed (i.d.). The state transition matrix $\mathbf{F}$ and the process noise covariance $\mathbf{Q}$ for the constant velocity motion are defined as \cite{1261132}
\begin{align}
\mathbf{F} &= \begin{bmatrix} 1 & \Delta_{T} \\ 0 & 1 \end{bmatrix} \quad \mathbf{Q} = \begin{bmatrix}
\frac{\Delta_{T}^{4}}{4}\sigma_{q}^{2} & \frac{\Delta_{T}^{3}}{2}\sigma_{q}^{2} \\
\frac{\Delta_{T}^{3}}{2}\sigma_{q}^{2} & \Delta_{T}^{2}\sigma_{q}^{2}
\end{bmatrix},
\end{align}
where $\Delta_{T}$ is the time step between epochs and $\sigma_{q}^{2}$ is the variance of the process noise that represents the uncertainty in the target's acceleration. 

We denote the positions of the Tx and Rx APs on the $x$-axis as $p_{\bar{\ell}}$ and $p_{\ell}$, respectively, and the user position as $(p_x,p_y)$ where $p_y$ is assumed to be fixed. For each pair $(\bar{\ell}\rightarrow \ell)$, define
\begin{subequations}
\begin{align}
\Delta_{\ell}[\kappa] &\triangleq p_x[\kappa] - p_{\ell}, \qquad
R_{\ell}[\kappa] \triangleq \sqrt{ \left(\Delta_{\ell}[\kappa]\right)^2 + p_y^2 },\\
\Delta_{\bar{\ell}}[\kappa] &\triangleq p_x[\kappa] - p_{\bar{\ell}}, \qquad
R_{\bar{\ell}}[\kappa] \triangleq \sqrt{ \left(\Delta_{\bar{\ell}}[\kappa]\right)^2 + p_y^2 }. \label{eq:distance difference and range from APs}
\end{align}
\end{subequations}
The observation relation is then given by
\begin{subequations}
\begin{equation}
\mathbf{h}(\mathbf{x}_\kappa) = \begin{bmatrix}
[\mathbf{h}(\mathbf{x}_\kappa)]_{\ell_1}^{\top},  
\ldots, 
[\mathbf{h}(\mathbf{x}_\kappa)]_{\ell_{L_\kappa}}^{\top}
\end{bmatrix},
\end{equation}
\begin{align}
&[\mathbf{h}(\mathbf{x}_\kappa)]_{\ell}=\Bigg[
\underbrace{\frac{R_{\bar{\ell}_1}[\kappa] + R_{\ell}[\kappa]}{c}}_{\tau_{\ell\bar{\ell}_1}(\mathbf{x}_\kappa)},
\ldots,
\underbrace{\frac{R_{\bar{\ell}_{|\mathcal{T}|}}[\kappa] + R_{\ell}[\kappa]}{c}}_{\tau_{\ell\bar{\ell}_{|\mathcal{T}|}}(\mathbf{x}_\kappa)},
\nonumber\\
&\underbrace{\frac{1}{\lambda}\!\left(\frac{\Delta_{\bar{\ell}_1}[\kappa]}{R_{\bar{\ell}_1}[\kappa]} + \frac{\Delta_{\ell}[\kappa]}{R_{\ell}[\kappa]}\right) v_x[\kappa]}_{\nu_{\ell\bar{\ell}_1}(\mathbf{x}_\kappa)},
\ldots\nonumber\\
&,\underbrace{\frac{1}{\lambda}\!\left(\frac{\Delta_{\bar{\ell}_{|\mathcal{T}|}}[\kappa]}{R_{\bar{\ell}_{|\mathcal{T}|}}[\kappa]} + \frac{\Delta_{\ell}[\kappa]}{R_{\ell}[\kappa]}\right) v_x[\kappa]}_{\nu_{\ell\bar{\ell}_{|\mathcal{T}|}}(\mathbf{x}_\kappa)},
\underbrace{\arctan\!\left(\frac{\Delta_{\ell}[\kappa]}{p_{y}}\right)}_{\theta_{\ell}(\mathbf{x}_\kappa)}
\Bigg],
\end{align}
\end{subequations}
where $c$ is the speed of light.
We set the measurement noise covariance as the ACM, $\mathbf{R}_{\kappa} = \boldsymbol{\mathcal{C}}_{\boldsymbol{\eta}_{\kappa}\boldsymbol{\theta}_{\kappa}}=\mathrm{blkdiag}[\boldsymbol{\mathcal{C}}_{\boldsymbol{\eta}_{\ell_1}\boldsymbol{\theta}_{\ell_1}},\ldots,\boldsymbol{\mathcal{C}}_{\boldsymbol{\eta}_{\ell_{\vert\mathcal{R}\vert}}\boldsymbol{\theta}_{\ell_{\vert\mathcal{R}\vert}}}]$. Here, $\boldsymbol{\eta}_{\kappa}$ stacks all $\{\tau_{\ell\bar{\ell}},\nu_{\ell\bar{\ell}}\}$ pairs over $\ell\in\mathcal{R}$ and $\bar{\ell}\in\mathcal{T}$, and $\boldsymbol{\theta}_{\kappa}$ stacks $\{\theta_{\ell}\}$ over $\ell$.
Given the state space model in \eqref{eq:state space}, the state $\mathbf{x}_{\kappa}$ can be recursively estimated using the EKF algorithm in Table \ref{table:EKF equations}. Note that the EKF recursions can predict the covariance even without new observations as $\mathbf{h}(\cdot)$ is linearized around the predicted state estimate.

\begin{table}[t!]
\centering
\caption{EKF equations used to estimate user state and uncertainty (covariance) of the estimate.}
\label{table:EKF equations}
\begin{tabular}{rl}
\hline
\multicolumn{2}{c}{\textbf{Time update (prediction step)}} \\
\hline
State prediction & $\hat{\mathbf{x}}_{\kappa|\kappa-1} = \mathbf{F}\mathbf{x}_{\kappa-1|\kappa-1}$ \\
Covariance prediction & $\mathbf{P}_{\kappa|\kappa-1} = \mathbf{F}\mathbf{P}_{\kappa-1|\kappa-1}\mathbf{F}^{\top} + \mathbf{Q}$ \\
\hline
\multicolumn{2}{c}{\textbf{Measurement update}$^{*}$} \\
\hline
EKF observation matrix & $\mathbf{H}_{\kappa} = \nabla \mathbf{h}(\mathbf{x})\bigg\rvert_{\mathbf{x}=\hat{\mathbf{x}}_{\kappa|\kappa-1}}$ \\
Innovation & $\boldsymbol{\varepsilon}_{\kappa} = \mathbf{z}_{\kappa} - h(\hat{\mathbf{x}}_{\kappa|\kappa-1})$ \\
Innovation covariance & $\mathbf{S}_{\kappa} = \mathbf{H}_{\kappa}\mathbf{P}_{\kappa|\kappa-1}\mathbf{H}_{\kappa}^{\top} + \mathbf{R}_{\kappa}$ \\
Kalman gain & $\mathbf{K}_{\kappa} = \mathbf{P}_{\kappa|\kappa-1}\mathbf{H}_{\kappa}^{\top}\mathbf{S}_{\kappa}^{-1}$ \\
State update & $\hat{\mathbf{x}}_{\kappa|\kappa} = \hat{\mathbf{x}}_{\kappa|\kappa-1} + \mathbf{K}_{\kappa}\boldsymbol{\varepsilon}_{\kappa}$ \\
Covariance update & $\mathbf{P}_{\kappa|\kappa} = (\mathbf{I} - \mathbf{K}_{\kappa}\mathbf{H}_{\kappa})\mathbf{P}_{\kappa|\kappa-1}$ \\
\hline
\end{tabular}\\
{\footnotesize {*} Measurement update is done only on sensing states.}
\end{table}

From the estimated user position, an estimate of the angle of arrival of the user observed at the origin can be calculated as 
\begin{align}
    \hat{\theta}^{ \text{\tiny EKF}}[\kappa] = g(\hat{p}_x[\kappa|\kappa^{*}])=\arctan\left(\frac{\hat{p}_x[\kappa|\kappa^{*}]}{p_{y}}\right).
\end{align}
The associated estimation error variance can be approximated as
\begin{align}
    \mathrm{var}(\tilde{\theta}^{\text{\tiny EKF}}[\kappa|\kappa^{*}]) 
    \simeq \nabla\mathbf{g}^{\top}\;\mathbf{P}_{\kappa|\kappa^{*}}\;\nabla\mathbf{g},
\end{align}
where $\nabla\mathbf{g}=\left[\dfrac{p_{y}}{p_{y}^{2}+\hat{p}_x[\kappa|\kappa^{*}]^{2}},\; 0\right]^{\top}$, which yields $\nabla\mathbf{g}^{\top}\mathbf{P}_{\kappa|\kappa^{*}}\nabla\mathbf{g} = \left(\frac{p_y}{p_y^2 + \hat{p}_x^2[\kappa|\kappa^{*}]}\right)^2
    [\mathbf{P}_{\kappa|\kappa^{*}}]_{1,1}$.
\section{Sensing Management}\label{sec: sensing management}

In this section, we propose methods for deciding when to perform sensing and which APs to use as Rx APs for sensing. The remaining APs will be assigned as Tx APs. The actions are determined based on the variance of the estimation error associated with the EKF-predicted angle of arrival of the user. These methods will be applied to each user. Therefore, we won't specify user indices except when necessary.

To implement the Rx AP selection, we introduce the matrix $\boldsymbol{\Omega}_{\kappa}=\mathrm{diag}(\omega_{1,\kappa},\ldots,\omega_{L_{\mathrm{T}},\kappa})\in \mathbb{Z}^{L_{\mathrm{T}}\times L_{\mathrm{T}}}_{2}$, where $\omega_{l,\kappa}\in\{0,1\} \ \text{for} \ l=1,\ldots, L_{\mathrm{T}}$. We determine that AP $l$ is an Rx AP if $\omega_{l,\kappa}=1$ and a Tx AP otherwise. The number of Rx APs serving in sensing at epoch $\kappa$ is $L_{\kappa}$.

\subsection{Sensing Time and Receive AP Selection} \label{subsec: Sensing Frequency Selection}
Herein, we consider the times when sensing signals need to be transmitted and perform tracking to maintain a certain performance in the angle estimate for each user. We want the estimated angles to be within the half-power beamwidth of the true angles to ensure high beamforming accuracy. Therefore, sensing waveforms are transmitted based on a comparison between the predicted variance of the angle estimation error and a threshold derived from the half-power beamwidth, as detailed in the appendix. 
As the user moves farther from an AP, the angular variation becomes increasingly negligible, causing the predicted estimation error variance to remain inherently small regardless of whether sensing is performed. Consequently, evaluating the sensing decision at a distant AP yields an unreliable criterion, as it would consistently favor skipping sensing even when the user's angle with respect to a nearer AP changes significantly. To address this, we evaluate the sensing decision based on the closest AP to each user, where angular variations are most pronounced and accurate tracking is most critical.

The closest AP is selected as $\ell_{k}^{*}=\min_\ell \Delta_{\ell k}[\kappa]$, where $\Delta_{\ell k}[\kappa]$ denotes the distance between AP $\ell$ and user $k$. We then determine the action for the next epoch for user $k$ as 
\begin{align}
    \resizebox{\linewidth}{!}{$\mathrm{Action}_{k}[\kappa+1] = \begin{cases*}
  \mathrm{Sensing}, & if  $ \mathrm{var}(\tilde{\theta}_{\ell_{k}^{*}k}^{\text{\tiny EKF}}[\kappa+1|\kappa_{k}^{*}]) \geq \gamma$ ,\\
  \mathrm{No \ Sensing},              & if $\mathrm{var}(\tilde{\theta}_{\ell_{k}^{*}k}^{\text{\tiny EKF}}[\kappa+1|\kappa_{k}^{*}]) < \gamma$ ,
  \end{cases*}$}
\end{align}
where $\kappa_k^{*}$ is the most recent epoch at which sensing was performed for user $k$. The predicted angle estimation error variance is $\mathrm{var}(\tilde{\theta}_{\ell_{k}^{*}k}^{\text{\tiny EKF}}[\kappa+1|\kappa_{k}^{*}]) \simeq \nabla\mathbf{g}_{\ell_{k}^{*}}^{\top}\mathbf{P}_{\kappa+1|\kappa_{k}^{*}}\nabla\mathbf{g}_{\ell_{k}^{*}}$, where $\nabla\mathbf{g}_{\ell_{k}^{*}}=\left[\dfrac{p_{y}}{p_{y}^{2}+\hat{\Delta}_{\ell_{k}^{*}}[\kappa|\kappa_k^{*}]^{2}},\; 0\right]^{\top}$ and $\hat{\Delta}_{\ell_{k}^{*}}[\kappa|\kappa_k^{*}] \triangleq \hat{p}_x[\kappa|\kappa_k^{*}] - p_{\ell_{k}^{*}}$.

We next address the selection of the receive APs so as to jointly support target tracking and downlink communication in the next epoch. In principle, the Rx/Tx partition could be obtained by solving a combinatorial optimization problem that minimizes the sum of the predicted angle of arrival estimation variances of all users undergoing sensing. However, such an exhaustive search quickly becomes prohibitive as the number of APs increases. This complexity is unnecessary because the communication utility is typically concentrated on only a few nearby APs. We therefore adopt a two-stage reduced-complexity design. 
First, we construct a small candidate set of dominant APs. Second, we optimize the Rx/Tx partition only over this set, while the APs outside the candidate set are kept as Tx APs.

\subsection{Utility-Based Candidate Set Construction}\label{subsec: Utility Order}

To quantify how much each AP contributes to the overall communication performance, 
we define for AP~$\ell$ the aggregate metric
\begin{equation}
    \xi_{\ell,\kappa}
    \triangleq
    \sum_{k \in \mathcal{K}_{\mathrm{C}}}
    \big\lvert \mathbf{h}_{\ell k,\kappa}^{\mathrm{H}}\,\mathbf{w}_{\mathrm{C},\ell k}\big\rvert^{2}.
    \label{eq:xi_def}
\end{equation}
The term $\xi_{\ell,\kappa}$ captures the total received signal power, summed over all communication users,
that is attributable to AP~$\ell$ with the current precoder.
Since both the channel vectors and the precoder are known at the CPU, the quantities $\{\xi_{\ell,\kappa}\}_{\ell=1}^{L_{T}}$ 
can be computed with negligible overhead at each epoch~$\kappa$.

The metric $\xi_{\ell,\kappa}$ can be seen as a per-AP communication utility. Therefore, we sort APs by $\xi_{\ell,\kappa}$ in descending order and denote the ordered indices as $\ell_1,\ell_2,\ldots,\ell_{L_T}$ such that
$\xi_{\ell_1,\kappa} \ge \xi_{\ell_2,\kappa} \ge \cdots \ge \xi_{\ell_{L_T},\kappa}$. We first keep all the $L_T$ APs in the active pool and decide how many of these should transmit using a cumulative stopping rule. We compute the running sum
$
C_{\ell_{\mathrm{c}}} \triangleq \sum_{i=1}^{\ell_{\mathrm{c}}} \xi_{\ell_i,\kappa}$ for $ \ell_{\mathrm{c}}=1,\ldots,L_T.
$
We then choose the smallest index $\ell_{\mathrm{c}}^{\star}$ such that
\begin{equation}
\ell_{\mathrm{c}}^{\star} = \min\Big\{ \ell_{\mathrm{c}} \in \{1,\ldots,L_T\} \;:\; C_{\ell_{\mathrm{c}}}/C_{L_{\mathrm{T}}} \ge \rho_{\mathrm{th}}  \Big\},
\end{equation}
with a design parameter $\rho_{\mathrm{th}} \in (0,1)$.
We assign the first $\ell_{\mathrm{c}}^{\star}$ APs as candidate set $
\mathcal{C}_\kappa \triangleq \{\ell_1,\ldots,\ell_{\ell_{\mathrm{c}}^{\star}}\}$.
$\mathcal{C}_{\kappa}$ typically contains only the few APs that dominate the communication utility.


It is important to emphasize that $\mathcal{C}_{\kappa}$ is only a reduced search set; it does not imply that all APs in $\mathcal{C}_{\kappa}$ are forced to be Tx or Rx APs. 

\subsection{Reduced-Complexity Receive AP Selection}

After constructing $\mathcal{C}_{\kappa}$, we determine the receive APs for epoch $\kappa+1$ by searching only over the APs in this set. The reduced search problem is then given by
\begin{subequations}
\begin{align}\label{eq: Angle estimation error variance with AP selection}
\underset{\breve{\boldsymbol{\Omega}}_{\kappa+1}}{\mathrm{minimize}}
\quad
&
\sum_{i \in \mathcal{K}_{\mathrm{S}}}
\nabla\mathbf{g}_{\ell_i^{*}}^{\top}
\mathbf{P}_{\kappa+1|\kappa_i^{*}}
\!\left(\breve{\boldsymbol{\Omega}}_{\kappa+1}\right)
\nabla\mathbf{g}_{\ell_i^{*}}
\\
\mathrm{subject\ to}
\quad
&
R_{i'}
\!\left(\breve{\boldsymbol{\Omega}}_{\kappa+1}\right)
\ge
R_{\mathrm{min}},
\qquad
i' \in \mathcal{K}_{\mathrm{C}},
\end{align}
\end{subequations}
where $R_{i'}$ is the rate of downlink communication for user $i'$ and $R_{\mathrm{min}}$ is the minimum rate that is allocated to each user. $R_{\mathrm{min}}$ is obtained by using the methods detailed in Section \ref{subsec:Minimum Rate for the Optimizaiton Problem}. $\mathbf{P}_{\kappa+1|\kappa_i^{*}}
\!\left(\breve{\boldsymbol{\Omega}}_{\kappa+1}\right)$ is the estimation error covariance matrix predicted with the given choice of Rx APs. The proposed method reduces the search space cardinality from $2^{L_{\mathrm{T}}}-2$ to $2^{\ell_{\mathrm{c}}^{\star}}-2$.

\section{Transmission Schemes} \label{sec: Transmission Schemes}

In Section \ref{sec: Downlink Signal Transmission}, we mentioned that we will consider three different transmission schemes within the proposed framework. In this section, we define their exact properties and then describe the proposed predictive precoding scheme.




\subsection{Sensing Waveforms}

In the SeS configuration, a fraction $0 \leq \mu_{\mathrm{c}}\leq 1$ of the OFDM subcarriers is reserved for sensing, i.e., $N_{\mathrm{c}}'=\mu_{\mathrm{c}}N_{\mathrm{c}}$. We allocate $N_{\mathrm{c}}'$ adjacent subcarriers to sensing, while the remaining subcarriers and the CP are dedicated to communication. The CP is excluded from radar range processing, as it creates a blind zone in the range cells. Furthermore, in the SeS case, the communication term in the sensing received signal \eqref{eq:single_snapshot_received_signal} vanishes.

In all cases, the per-AP transmit power is split so that the fraction $\mu_{\mathrm{p}}$ is reserved for sensing while the fraction $1-\mu_{\mathrm{p}}$ is used for communication. Assuming a uniform power allocation across subcarriers, the sensing power budget in SeS scales linearly with $\mu_{\mathrm{c}}$, since only a $\mu_{\mathrm{c}}$-portion of subcarriers carries the sensing power. Therefore, for the SeS case, the power fraction is equal to the subcarrier fraction $\mu_{\mathrm{p}}=\mu_{\mathrm{c}}.$


In SS-SW, one must decide which communication data streams to reuse for sensing so that the communication and sensing beams reinforce each other rather than interfere. To limit the interference imposed on downlink communication users, we employ a set-allocation mechanism.
To select an appropriate user $k$ for a sensing target $i$, we introduce a heuristic communication-aware radar alignment score. This score quantifies how well the sensing beam for target $i$ aligns with user $k$ while accounting for the interference it generates for other active users
\begin{equation}
\mathcal{J}_{k,i}\triangleq \frac{\big\lvert \mathbf{h}_k^{\top}\,\mathbf{w}_{\mathrm{S},i}\big\rvert^{2}}{ \sum_{j\in\mathcal{K}_{\mathrm{C}},\, j\neq k} \big\lvert \mathbf{h}_j^{\top}\,\mathbf{w}_{\mathrm{S},i}\big\rvert^{2}}. \label{eq:CARA}
\end{equation}
The user is chosen as the candidate partner for the target $i$ based on 
\begin{equation}
\hat{k}(i)=\arg\max_{k\in\mathcal{K}_{\mathrm{C}}}\mathcal{J}_{k,i}.
\end{equation}
This allocation is done for each target. We denote the collection of all targets that share the data signal with user $k$ by $\mathcal{D}_{k}$.

Note that the overall precoder then becomes a superposition of the communication precoder and the matched sensing precoders as the transmitted signals are common for the matched target-user pair
\begin{equation}
\mathbf{w}_{\mathrm{tot},k}
=\mathbf{w}_{\mathrm{C},k}
+\sum_{i\in\mathcal{D}_{k}}\mathbf{w}_{\mathrm{S},i}. 
\label{eq:sssw_wtot}
\end{equation}


In the SS-SeW and SeS cases, the sensing signals do not depend on the user index $i$, but depend on the Tx AP index. Sensing codes are chosen to be deterministic, low-correlation sequences embedded in the subcarriers. Furthermore, a code assignment per Tx AP is also considered to ensure low inter-AP cross-correlation.

\subsection{Predictive MMSE Precoding} \label{Predictive MMSE Precoder}
Next, we explain the proposed MMSE precoder, formed from the predicted states obtained from the tracking filter incorporating the movement model.


We note that we estimate the angle and the channel coefficient of the LOS component of the user channels, but the NLOS components are not estimated. We first obtain the pathloss coefficients from the range estimates as $\hat{\beta}_{\bar{l}i,\kappa}=\frac{\lambda^2}{(4\pi\hat{R}_{\bar{l}i,\kappa})^2}$, where the range estimate is found with \eqref{eq:distance difference and range from APs} using the position estimates. Then we define the channel estimates as $\hat{\bar{\mathbf{h}}}_{\bar{l}i}[\kappa\vert\kappa^{*}_{i}]=\sqrt{\hat{\beta}_{\bar{l}i,\kappa}}\mathbf{a}(\hat{\theta}_{\bar{l}i}[\kappa\vert\kappa^{*}_{i}])$, we stack them as $\hat{\bar{\mathbf{h}}}_{i}[\kappa\vert\kappa^{*}_{i}]=\Big[\hat{\bar{\mathbf{h}}}_{\bar{\ell}_{1}i}[\kappa\vert\kappa^{*}_{i}],\ldots,\hat{\bar{\mathbf{h}}}_{\bar{\ell}_{|\mathcal{T}|}i}[\kappa\vert\kappa^{*}_{i}]\Big]^{\top}\in\mathbb{C}^{|\mathcal{T}|N\times1}$. Using these, we obtain the MMSE precoder for user $k$ at epoch $\kappa$ as
\begin{align}
    &\bar{\mathbf{w}}_{k}[\kappa\vert\kappa^{*}_{k}]=q_{k,\kappa}\Bigg[ \sum_{i=1}^{K} q_{i,\kappa}  
\frac{K_{\mathrm{R}}}{1+K_{\mathrm{R}}} \Bigg(   \hat{\bar{\mathbf{h}}}_{i}[\kappa\vert\kappa^{*}_{i}]\hat{\bar{\mathbf{h}}}_{i}^{\top}[\kappa\vert\kappa^{*}_{i}]   + \nonumber\\& \mathbf{C}_{\tilde{\bar{\mathbf{h}}}_{i}}[\kappa\vert\kappa^{*}_{i}] \Bigg)  +
\frac{1}{1+K_{\mathrm{R}}}\,
\mathbf{R}_{{\rm NLOS},i}
+  \sigma_{n}^2\mathbf{I}_{|\mathcal{T}|N}  \Bigg] ^{-1} \nonumber\\& \times \sqrt{\frac{K_{\mathrm{R}}}{1+K_{\mathrm{R}}}}\hat{\bar{\mathbf{h}}}_{k}[\kappa\vert\kappa^{*}_{k}],
\end{align}
where $\mathbf{C}_{\tilde{\bar{\mathbf{h}}}_{i}}[\kappa\vert\kappa^{*}_{i}]=\mathbb{E}\left\{\tilde{\bar{\mathbf{h}}}_{i}[\kappa\vert\kappa^{*}_{i}] \tilde{\bar{\mathbf{h}}}_{i}^{\mathrm{H}}[\kappa\vert\kappa^{*}_{i}] \right\}$ captures the angle estimation error for user $i$, $\mathbf{R}_{{\rm NLOS},i}
$ captures the covariance of the NLOS part of the channel, and $q_{i,\kappa}$ is the transmit power of user $i$ normalized by the noise power in the virtual uplink.

A power fraction is employed to the precoder with $\bar{\mathbf{w}}_{\mathrm{S},k}[\kappa\vert\kappa^{*}_{k}]=\sqrt{\mu_{\mathrm{p}}}\bar{\mathbf{w}}_{k}[\kappa\vert\kappa^{*}_{k}]$ for the sensing precoder and $\bar{\mathbf{w}}_{\mathrm{C},k}[\kappa\vert\kappa^{*}_{k}]=\sqrt{(1-\mu_{\mathrm{p}})}\bar{\mathbf{w}}_{k}[\kappa\vert\kappa^{*}_{k}]$ for the communication precoder. The precoder can be obtained along similar lines to \cite{demir2021foundations}, using the predicted information as the channel state information.
The per-AP power constraints, defined in Section \ref{sec: Downlink Signal Transmission}, must be satisfied. Hence, we normalize the precoders to user $k$ as 
\begin{equation}
\mathbf{w}_{k}[\kappa\vert\kappa^{*}_{k}] = \sqrt{\rho_{k,\kappa}}\frac{\bar{\mathbf{w}}_{k}[\kappa\vert\kappa^{*}_{k}]}{\sqrt{ \sum_{\bar{l}\in\mathcal{T}} \mathbb{E} \left\{ \Vert\bar{\mathbf{w}}_{\bar{l}k}[\kappa\vert\kappa^{*}_{k}]\Vert^{2} \right\}}},
\end{equation}
where $\bar{\mathbf{w}}_{\bar{l}k}[\kappa\vert\kappa^{*}_{k}]$ is the precoder from AP $\bar{l}$ to user $k$ and $\rho_{k,\kappa}$ determines how much power is allocated to this user.
In the numerical results, we consider
\begin{equation}
\rho_{k,\kappa}=\rho_{d}\frac{\sqrt{\sum_{\bar{l}\in\mathcal{T}}\hat{\beta}_{\bar{l}k,\kappa}}}{\max\limits_{\bar{l}}\sum_{i\in\mathcal{K}_{\mathrm{C}}}\sqrt{\sum_{\bar{l}\in\mathcal{T}}\hat{\beta}_{\bar{l}i,\kappa}}}, 
\end{equation}
which slightly favors users with better channel conditions. 



\section{Downlink Achievable SE Results} \label{sec: SE Results}

In this section, we state the SE expressions for the three transmission schemes. By assuming perfect CSI at the users, we obtain an upper bound on the achievable spectral efficiency. 

\subsection{Spectral Efficiency with Shared Subcarriers}

With perfect CSI at the user, an achievable ergodic SE at user $k$ is \cite{demir2021foundations} 
\begin{equation} \label{Perfect_CSI_SE}
    \mathrm{SE}_{k} =  \mathbb{E}\left\{  \log_{2} \left(  1 +  \mathrm{SINR}_{k}  \right) \right\}.
\end{equation}
The communication SINR of user $k$ at a considered coherence block can be expressed as
\begin{equation}\label{eq: SINR with Shared Subcarriers}
\mathrm{SINR}_{k} =  \,\frac{\vert b_{kk}\vert^{2}}{\sum_{i\in\mathcal{K}_{C}, i\neq k}  \vert b_{ki}\vert^{2}+\sum_{i\in\mathcal{K}_{s}}\vert\tilde{b}_{ki}\vert^{2}+\sigma_n^2},
\end{equation}
where $b_{ki}=\mathbf{h}^{\top}_{k}[(\mathbf{I}_{L_{\mathrm{T}}}-\boldsymbol{\Omega}_{})\otimes\mathbf{I}_{N}]\mathbf{w}_{i}$ is the effective channel between the channel of user $k$ and the precoder of user $i$. The operation $(\mathbf{I}_{L_{\mathrm{T}}}-\boldsymbol{\Omega}_{})\otimes\mathbf{I}_{N}$ chooses the APs left for transmission and widens the matrix to fit the dimension of the precoder. $\tilde{b}_{ki}=\mathbf{h}^{\top}_{k}[(\mathbf{I}_{L_{\mathrm{T}}}-\boldsymbol{\Omega}_{})\otimes\mathbf{I}_{N}]\mathbf{w}_{\mathbf{S},i}$ is the effective channel between the user $k$ and the sensing precoder towards sensing target $i$. $\mathbf{h}_{k}=[\mathbf{h}_{1k},\ldots,\mathbf{h}_{L_{\mathrm{T}}k}]^{\top}$ is the stacked channel and $\mathbf{w}_{k}=\left[\mathbf{w}_{1k},\ldots,\mathbf{w}_{L_{\mathrm{T}}k}\right]^{\top}$ is the stacked precoder between all APs and the user $k$.
$\sigma_n^2$ is the noise variance.

Note that, for the SS-SW case, the effective channel contains the total precoder towards each user $k$ and is equal to $ b_{kk}  + \sum_{i\in \mathcal{D}_{k}} \tilde{b}_{ki}$. Moreover, the sensing interference term decreases and equals to $\sum_{i\in\mathcal{K}_{s}\backslash\mathcal{D}_{i}}\vert\tilde{b}_{ki}\vert^{2}$.



\subsection{Spectral Efficiency with Separate Subcarriers}
Next, we consider the case of separate subcarriers. We note that $1-\mu_{\mathrm{c}}$ portion of the frequency resources is allocated for the communication. With perfect CSI at the users, an achievable SE at user $k$ is 
\begin{equation} \label{Perfect_CSI_SE_SeS}
    \mathrm{SE}_{k} =  (1-\mu_{\mathrm{c}})\mathbb{E}\left\{  \log_{2} \left(  1 +  \mathrm{SINR}_{k}  \right) \right\}.
\end{equation}
The communication SINR of user $k$ at a considered coherence block can be expressed as
\begin{equation}\label{eq: SINR with separate subcarriers}
\mathrm{SINR}_{k} =  \,\frac{\vert b_{kk}\vert^{2}}{\sum_{i\in\mathcal{K}_{C}, i\neq k}  \vert b_{ki}\vert^{2}+\sigma_n^2},
\end{equation}
where $b_{ki}=\mathbf{h}^{\top}_{k}[(\mathbf{I}_{L_{\mathrm{T}}}-\boldsymbol{\Omega}_{})\otimes\mathbf{I}_{N}]\mathbf{w}_{i}$ is the effective channel between the channel of user $k$ and the precoder of user $i$. In this case, there is no interference between sensing waveforms and communication users, as the frequency resources are separated. We note that the main differences between the two cases are the interference terms and the subcarrier allocation.


\subsection{Minimum Rate for the Optimization Problem}\label{subsec:Minimum Rate for the Optimizaiton Problem}

Here, we explain how the minimum rate in the optimization problem \eqref{eq: Angle estimation error variance with AP selection} is obtained. We evaluate the instantaneous rates $R_{i'}=\log_{2} \left(  1 +  \mathrm{SINR}_{i'}(\mathcal{T}_{\kappa})  \right)$ for each user by setting $\mathcal{T}_{\kappa}=\mathcal{C}_{\kappa}$ and setting the rest as Rx APs. The SINR relations are found using the \eqref{eq: SINR with separate subcarriers} and \eqref{eq: SINR with Shared Subcarriers} for the SeS and SS cases, respectively. Then we find the minimum rate as 
\begin{align}
    R_{\mathrm{min}}=\min_{i'\in \mathcal{K}_{C}} R_{i'}.
\end{align}


\section{Simulation Setting and Results}


In this section, we evaluate the performance of the proposed predictive beamforming scheme in a cell-free massive MIMO system similar to the one shown in Fig.~\ref{fig:APs and car system}. In our scenario, we assume the APs are located equidistantly on a horizontal line along the road. The vehicles are assumed to move in the positive direction, following a constant-velocity motion model with stochastic perturbations captured by the EKF process noise.
Unless otherwise noted, the system parameters are as given in Table~\ref{tab:sim_params}.
The initial uncertainty in the position and velocity of the user is characterized by the covariance matrix $\mathbf{P}_{0|0}=\mathrm{diag}(1000,25)$. The sensing-time selection threshold is $\gamma = 14.9^{\circ}$ for $N=8$ antennas. We assume that the RCS fluctuates according to the Swerling I model with a mean value $ 5\ \textrm{m}^ {2}$. Note that we assume we do not have accurate knowledge of the target's initial locations. Moreover, we allow the targets to accelerate and decelerate significantly. These parameter settings are deliberately chosen to highlight the robustness of the proposed method under poor prior information.

\begin{table}[t!]
\centering
\caption{SIMULATION PARAMETERS}
\label{tab:sim_params}
\begin{tblr}{|l|c|}
\hline
Parameter & Value \\
\hline
Number of AP antennas $(N)$ & 8 \\
\hline
Number of APs ($L$) &  20 \\
\hline
Time difference  & $\triangle_T=0.01$ s \\
\hline
Acceleration noise parameter  & $ \sigma_q^{2}=1$ \\
\hline
Position of the $l$th AP $(x,y)$ & $((2000/L)l,0)$\\
\hline
Vehicles' start position $(x,y)$ & $(\mathcal{U}[0,2000],40)$ \\
\hline
Carrier frequency $(f_c)$ & 30 GHz \\
\hline
Number of epochs & 5000 \\
\hline
Velocity $(v)$ & 20 m/s \\
\hline
Noise variance $(\sigma_n^2)$ & $-95$ dBm \\
\hline
Transmit power per AP $(\rho_{d})$ & 200 mW \\
\hline
\end{tblr}
\end{table}

We assume that $\mu_{p}=0.05$, $\mu_{c}=0.05$ and $\rho_{\mathrm{th}}=0.95$ in all the scenarios. Moreover, we assume $K=4$ users in the system unless otherwise stated. Please note that, although there are $20$ APs, the APs in the close vicinity of the user contribute most to both sensing and communication. Figures showing the temporal evolution of the SEs indicate the SE that the system can support at a particular epoch if a communication request is made. The SE results are obtained by taking a 100-epoch-centered moving average of the instantaneous rates.
We assume that $5$ epochs correspond to a coherence block, and therefore, we take the mean over $20$ coherence blocks to obtain the SE results. In practice, if the sensing and communication epochs coincide, the communication signals can be transmitted simultaneously with the sensing signals using a subcarrier separation as in the SeS method, but applied to the same user. In the simulations, we assume that communication requests occur only during epochs when there is no sensing need. Therefore, SE curves report the achievable rate during epochs allocated to communication (i.e., epochs in which sensing is not scheduled). We assume that Barker-13 coded pulse trains are used in the SS-SeW and SeS cases. 



\subsection{Evolution of Angle Estimation Variance with Sensing Decisions}
\begin{figure}[t!]
\centering
  \includegraphics[width=\linewidth]{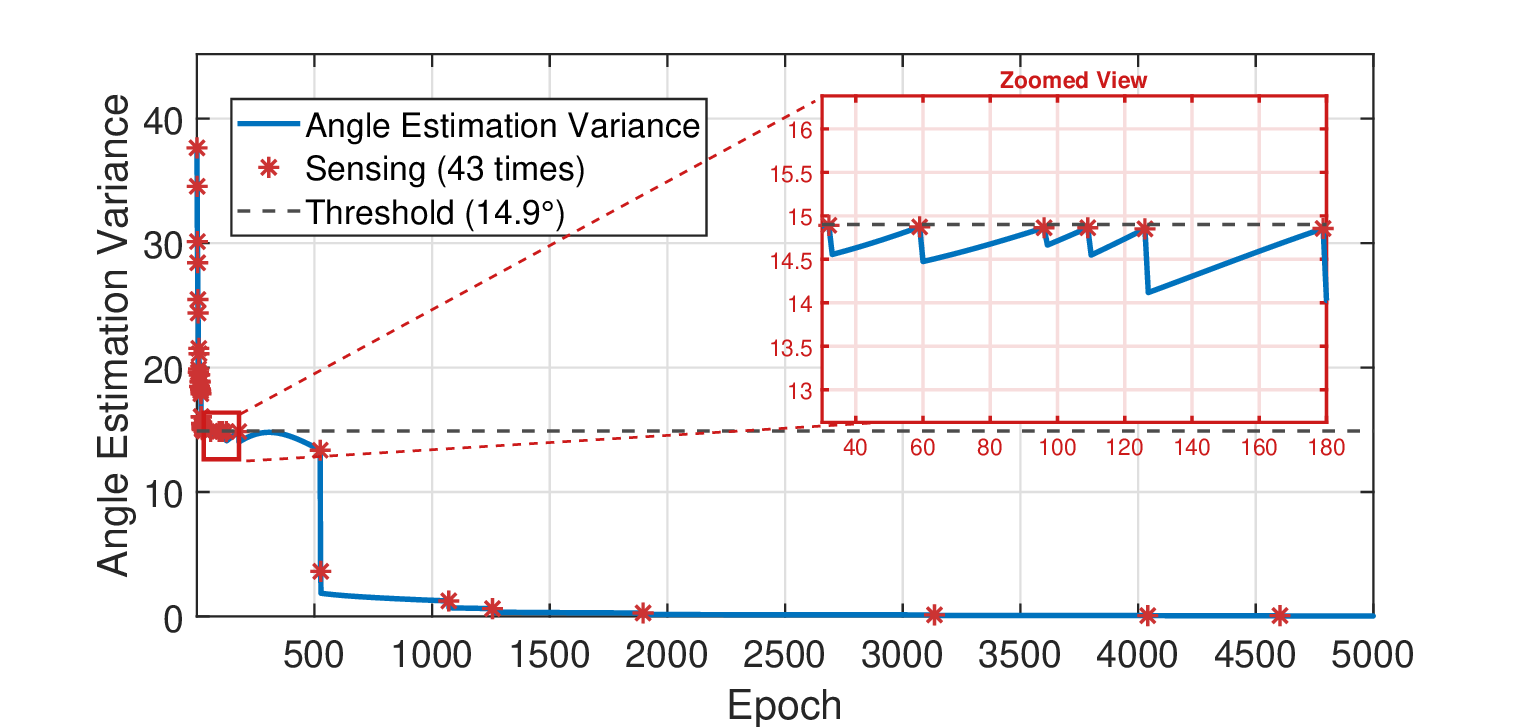}\vspace{-3mm}
    \caption{The temporal behavior of the predicted angle estimation variance and sensing decisions with optimal Rx AP selection.}\vspace{-3mm}
    \label{fig:angle_var}
\end{figure}

We first investigate the impact of sensing decisions on the angle estimation variance. Fig.~\ref{fig:angle_var} illustrates the temporal behavior of the predicted angle estimation variance with respect to the first AP with the proposed Rx AP selection. 
We consider a user initialized at the origin to clearly illustrate the temporal variations in this figure.
As described in Section~\ref{sec: sensing management}, the proposed sensing management method triggers a sensing decision whenever the predicted angle estimation variance reaches the predefined threshold, as shown in the zoomed-in portion. During the initial sensing epochs, the variance relative to the first AP occasionally exceeds the threshold, triggering sensing decisions. As the user moves away from the first AP and toward subsequent APs, the angle estimation variance associated with the first AP remains consistently below the threshold. Subsequent sensing decisions are therefore governed by the remaining APs, which is closest to the user.
Fig.~\ref{fig:angle_var} shows that the angle estimation variance drops significantly when there is a sensing decision in the former epoch. We can see that there isn't a need to frequently transmit sensing signals while maintaining highly accurate angle estimates, except at the beginning, since the initial uncertainty was assumed to be high. 
We can see that sensing is needed roughly $1\%$ of the time. 

\subsection{Performance of the Transmission Schemes} \label{subsec: Fig 2}

\begin{figure}[t!]
\centering
  \includegraphics[width=\linewidth]{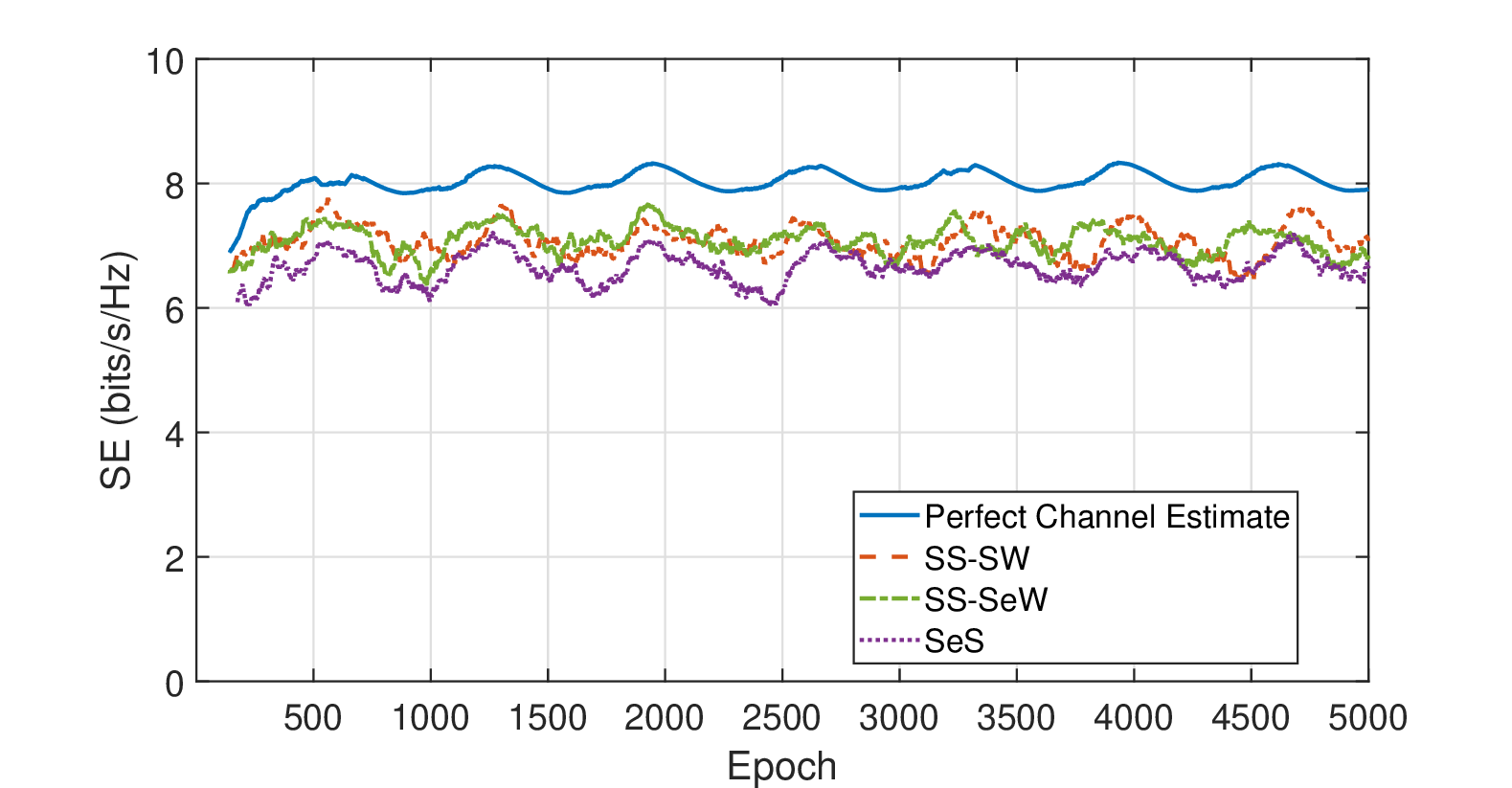}\vspace{-3mm}
    \caption{The temporal behavior of the SEs achieved with the three transmission schemes and the perfect channel knowledge at the APs.}\vspace{-3mm}
    \label{fig:threecases}
\end{figure}

Next, we investigate the behavior of different transmission methods. Fig.~\ref{fig:threecases} shows the SE of one of the four uniformly randomly initiated users with the three transmission schemes as a function of the epochs. We consider the perfect CSI case as a benchmark. We assume that perfect CSI is achieved by the sensing and tracking steps.
From this figure, we can observe that all methods remain within a small gap from the perfect CSI case. SS-SeW achieves the highest SE in most epochs by using a dedicated sensing waveform, yielding the most accurate parameter estimates and, consequently, the most reliable beamforming. SS-SW performs close to SS-SeW, but its random sensing waveform leads to noisier estimates and occasional SE degradation. SeS attains slightly lower SE because only (5$\%$) of the subcarriers are allocated to sensing; nevertheless, the gap remains limited, indicating that a small sensing resource budget can be sufficient when dedicated radar-like pulses are used.\footnote{If random waveforms were instead used with a low number of subcarriers, estimation performance would be significantly lower, leading to very poor SE performance. Therefore, the plot for this variant is omitted. This implies that random waveforms work well only when a high number of subcarriers are allocated for sensing.}

The small SE variations result from divergence in the angle estimate until the sensing decision is triggered. Larger fluctuations, observed even with perfect channel estimates, are caused by changes in distance to the APs. The SE is otherwise relatively uniform across epochs due to joint AP operation. The gap between the benchmark and the proposed method is primarily due to the unknown realizations of the channel’s NLOS component.

\subsection{Impact of the Number of Users}
\begin{figure}[t!]
\centering
  \includegraphics[width=\linewidth]{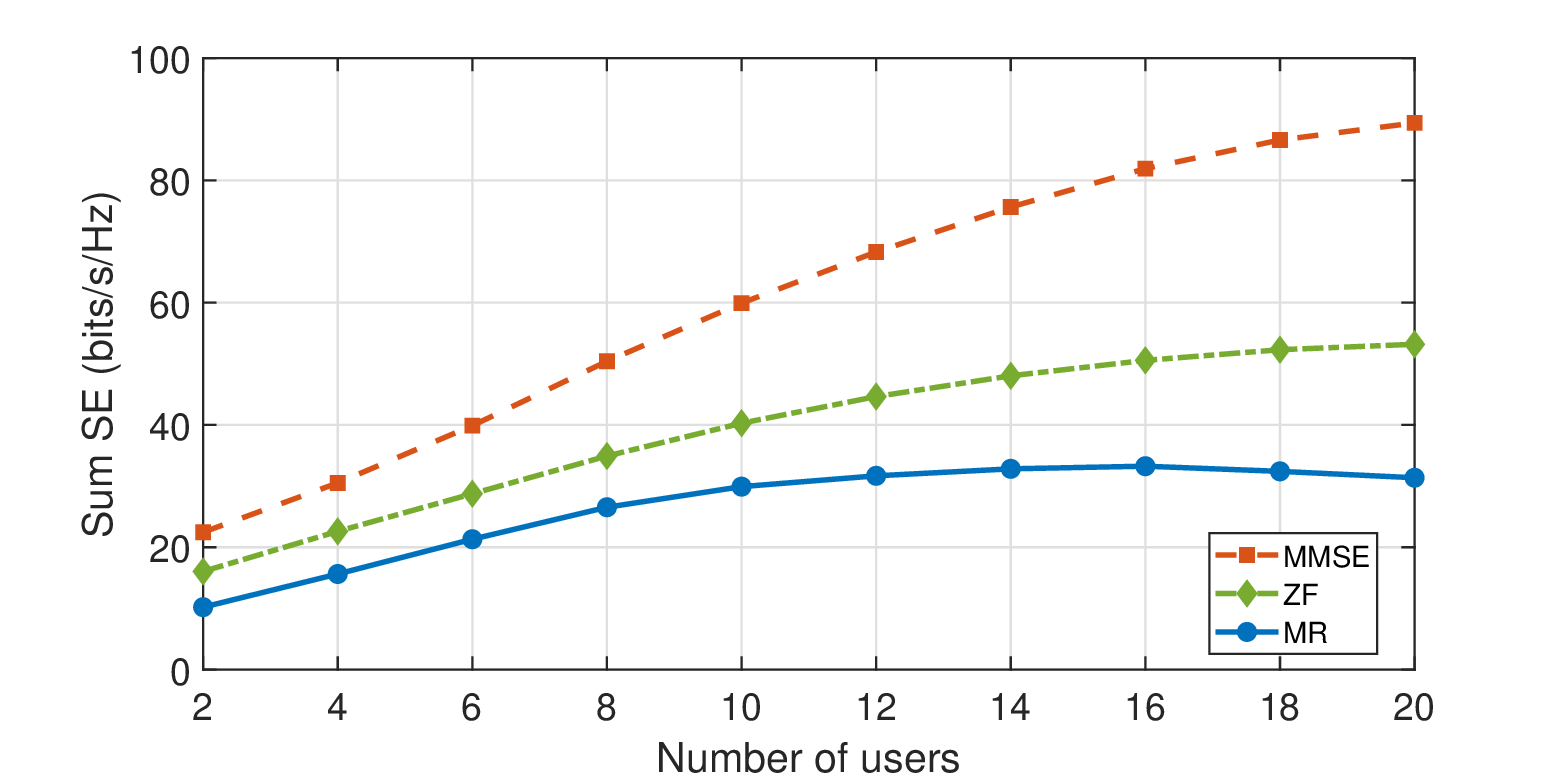}\vspace{-3mm}
    \caption{Achievable sum SE as a function of the number of users with MMSE, ZF and MR precoders.}\vspace{-3mm}
    \label{fig:fig3}
\end{figure}

In this subsection, we investigate the impact of the number of users on the SE results.
Fig.~\ref{fig:fig3} shows the variation of the sum SE results with the number of users. From this figure, we can see that the proposed MMSE precoder performs significantly better than both the ZF and the MR precoders. 


At large $K$, the MR precoder becomes interference-limited, leading to a reduction in the sum SE. The MMSE precoder mitigates this trend by actively suppressing multi-user interference, hence maintaining a higher sum SE. Another reason is that, with $20$ users, many are located in close proximity, which affects parameter estimation quality. The MMSE precoder handles this effect as well. With ZF precoder, we don't see a decrease as the number of users increases. However, it doesn't perform as well as the MMSE precoder as it is unable to suppress the NLOS part of the channel, due to lack of channel estimates.

\subsection{Impact of the Number of AP Antennas}
\begin{figure}[t!]
\centering
  \includegraphics[width=\linewidth]{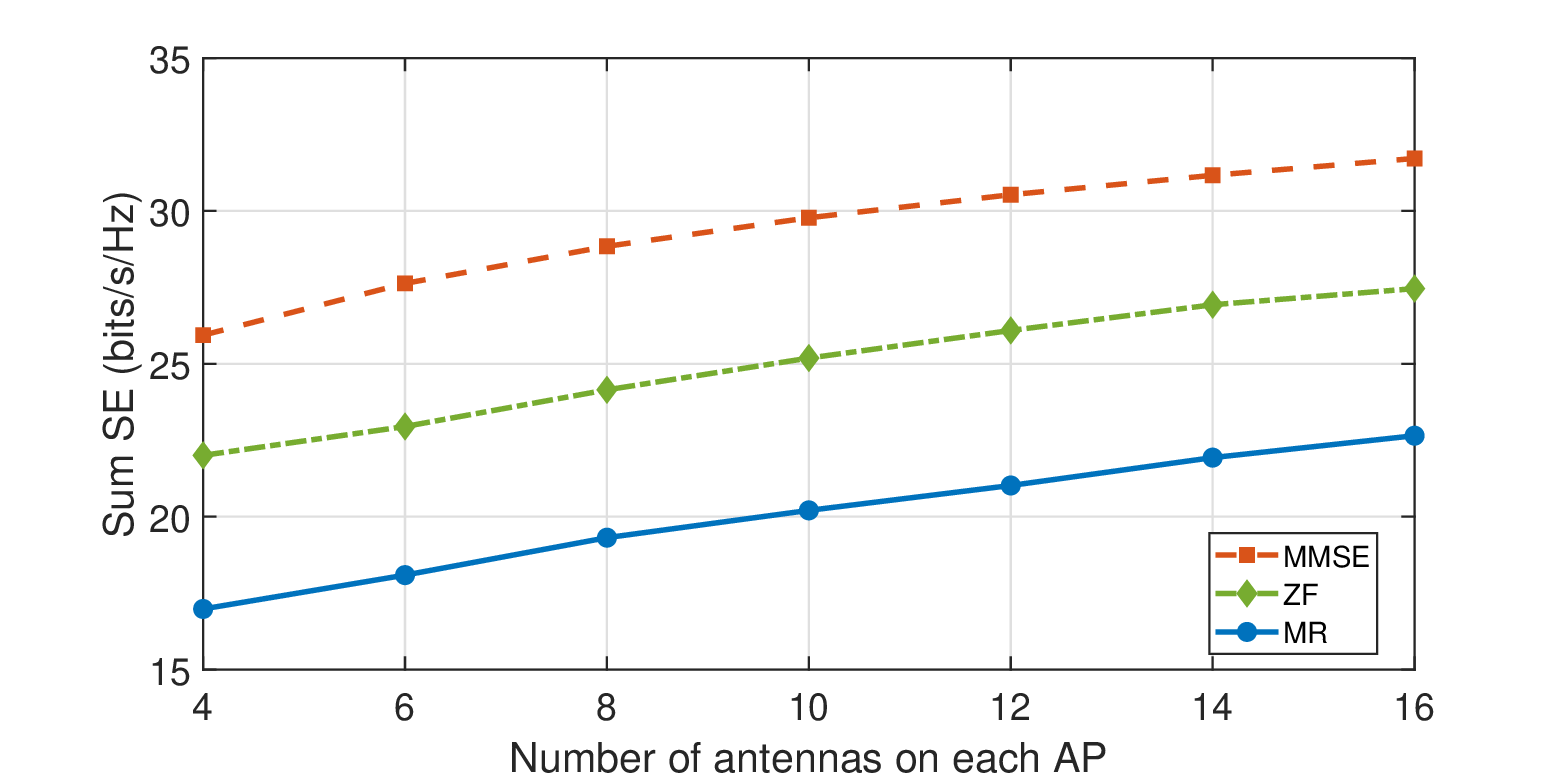}\vspace{-3mm}
    \caption{Achievable sum SE as a function of the number of AP antennas with MMSE, ZF and MR precoders.}\vspace{-3mm}
    \label{fig:fig4}
\end{figure}

Next, we investigate the impact that the number of AP antennas has on the SEs. Fig.~\ref{fig:fig4} shows the sum SE when $N$ varies from $4$ to $16$. 
We note that the sensing-time selection threshold increases with the number of antennas. The threshold values $\gamma=\{60.465, 26.639, 14.938, 9.545, 6.624, 4.864, 3.723\}$ correspond to the antenna numbers $N=\{4, 6, 8, 10, 12, 14, 16\}$. We observe that the proposed MMSE precoder outperforms both the ZF and MR precoders throughout the range of antenna values. This performance gain arises from the inherent capability of the MMSE precoder to jointly mitigate inter-user interference and account for the estimation errors.
We note that as the threshold value decreases, the number of sensing iterations naturally increases. However, as the number of antennas increases, we still observe an increase in the sum SE.


\subsection{Comparison to Prior Works}
\begin{figure}[t!]
\centering
  \includegraphics[width=\linewidth]{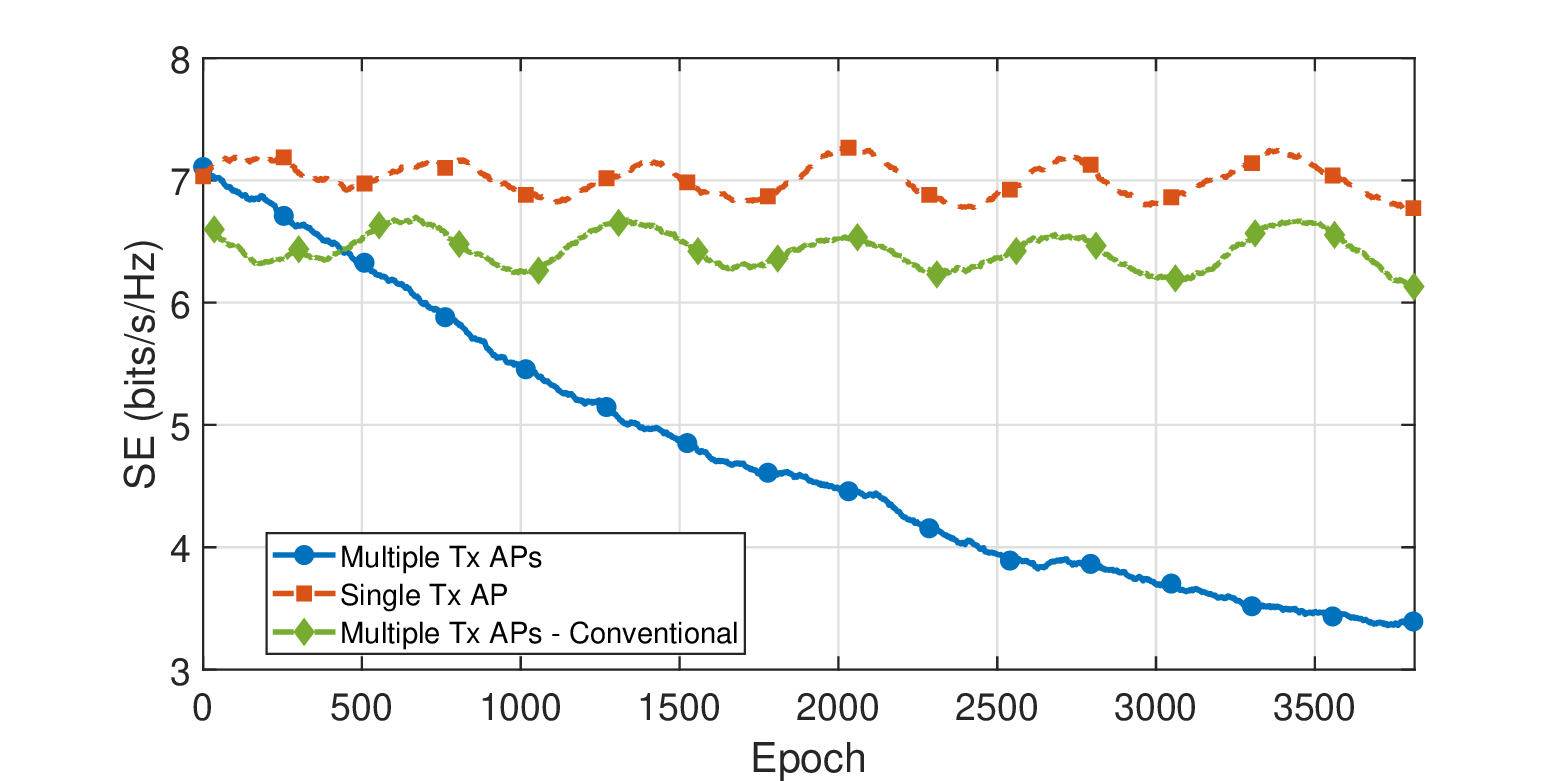}\vspace{-3mm}
    \caption{The temporal behavior of SEs with multiple Tx APs with the proposed and conventional method and a single Tx AP with conventional method and $N=40$ antennas.  }\vspace{-3mm}
    \label{fig:fig5}
\end{figure}

Here, we compare the proposed method with prior works \cite{akçalı2025predictivebeamformingdistributedmimo,9171304}. Fig.~\ref{fig:fig5} shows the evolution of the SE results for a user with the proposed method, a single-transmitter with $N=40$ antennas, and the cell-free massive MIMO case with the conventional approach. By conventional approach, we mean the approach described in Section~\ref{sec:Communication and Sensing System Model}. For the conventional case, as the proposed sensing management isn't used, a continuous power allocation between sensing and communication is considered. $\mu_{p}$ portion of the transmit power $\rho_{d}$ is continuously used for sensing, and the rest is reserved for the communication signal transmission. We plot the $4000$ epochs in this figure to better show the fluctuations in the proposed method compared to the drop in the single-Tx AP case. Furthermore, we omit the initial sensing epochs in the single-Tx AP case. This is done as we intentionally assume poor initial knowledge of the states in all our simulations to show the performance of the proposed method.



We see that the proposed sensing management maintains a stable communication performance over time. This stems from the broader geographical coverage due to distributed APs and accurate tracking. Fig.~\ref{fig:angle_var} and Fig.~\ref{fig:fig5} together show that channel estimation isn't needed frequently, or reactively when there is data to transmit. Instead, one can track the UE and perform sensing when the angle estimation error variance increases, and use the tracked angle for communication when necessary, as proposed.

\subsection{SE Results of Multiple Closely Initiated Users}
\begin{figure}[t!]
\centering
  \includegraphics[width=\linewidth]{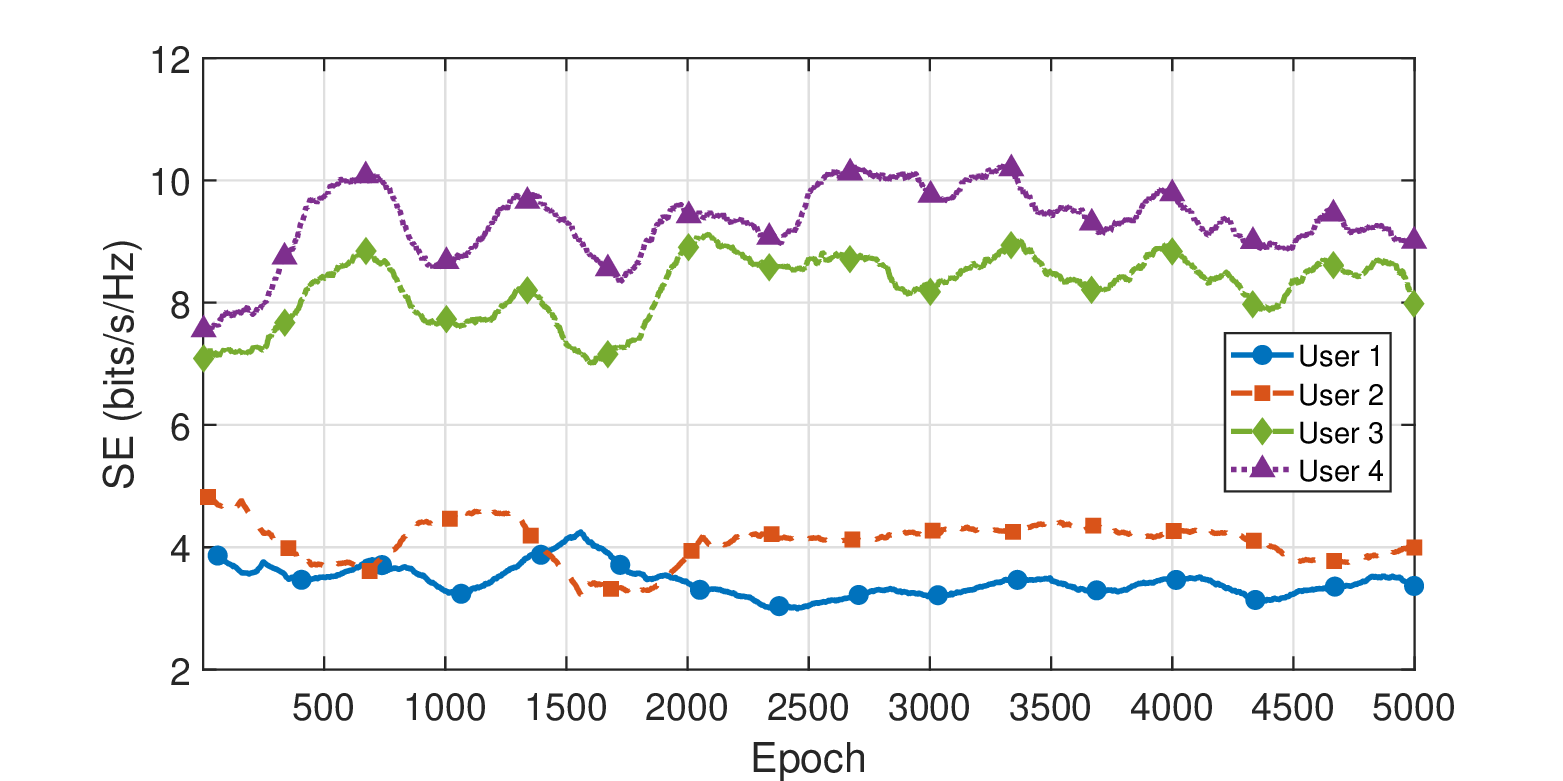}\vspace{-3mm}
    \caption{The temporal behavior of SEs with the proposed method and SS-SeW with $4$ closely initiated users.}\vspace{-3mm}
    \label{fig:fig6}
\end{figure}

Next, we study the SE when multiple users exist in a close proximity in both range and velocity. This case is particularly important for real-world applications, as users are rarely separated by significant distances.
Fig.~\ref{fig:fig6} shows the evolution of the SE of 4 users with the proposed method. We assume that the four users are initially at the positions $p_x=\{0, 10, 100, 300\}$m. We can see that all users have high SEs, even though they are located close to each other. This happens as the EKF is highly capable of processing the received parameter estimates.

This setting is particularly challenging from an estimation perspective, as the reflected signals from closely spaced users are not well separated.
Despite this inherent coupling, Fig.~\ref{fig:fig6} indicates that all users maintain high and relatively uniform SE throughout their trajectories. We note that Users 2 and 3 are initialized with only $10$\,m separation; consequently, a larger number of sensing updates is required before sufficiently accurate parameter estimates are obtained and the SE stabilizes. Moreover, we observe a see-saw pattern between these two nearby users: when one attains a higher SE, the other tends to experience a reduction. This behavior is mainly due to mutual interference from the other users’ precoders under limited separability. This effect can be solved through power allocation optimization.

\subsection{Tracking of Position and Angle of Users}
\begin{figure}[t!]
\centering
  \includegraphics[width=\linewidth]{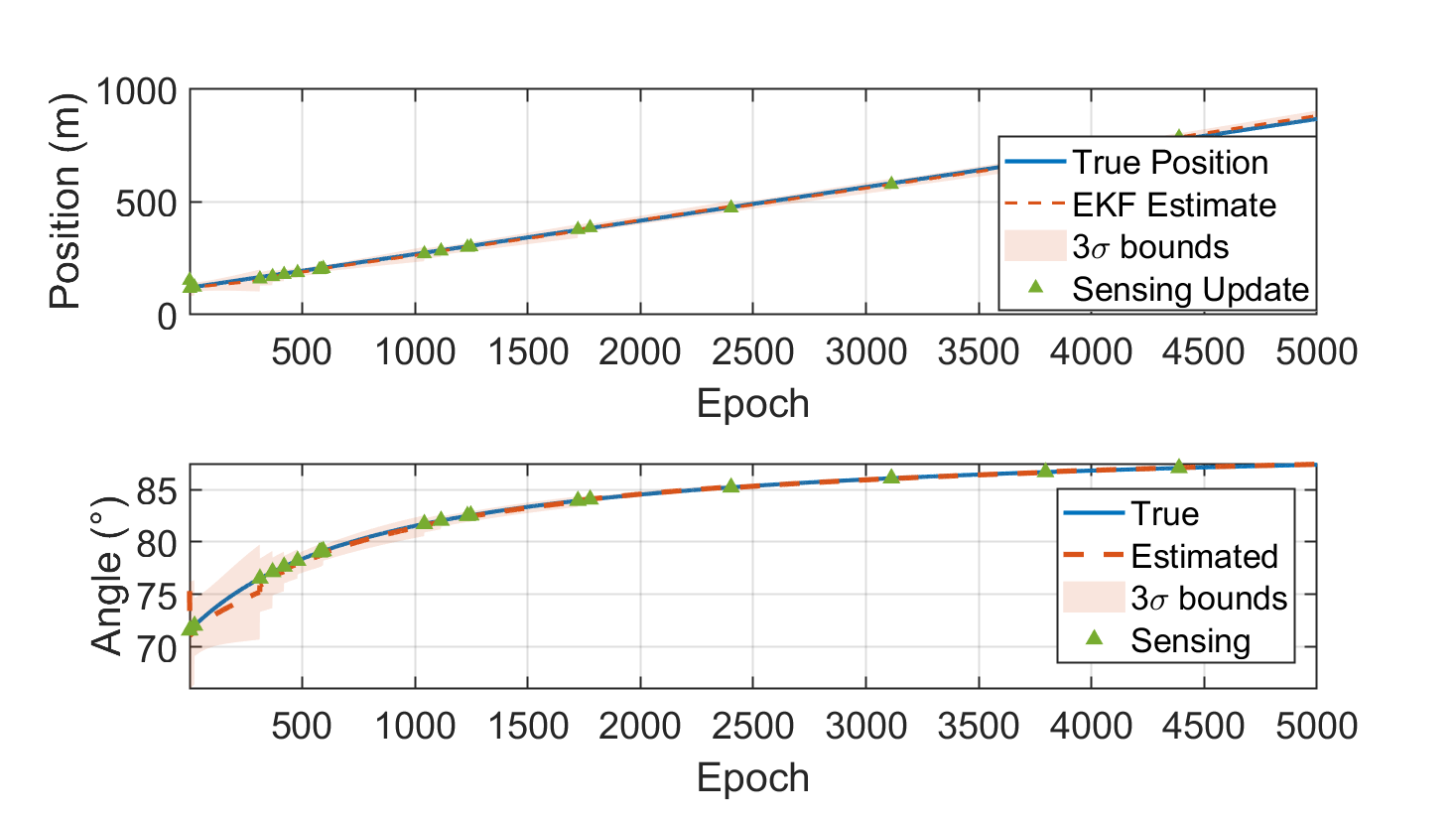}\vspace{-3mm}
    \caption{Tracking of the position and angle of a user with respect to the first AP.}\vspace{-3mm}
    \label{fig:fig7}
\end{figure}

Finally, we investigate the tracking of position and angle with the proposed method.
Fig.~\ref{fig:fig7} shows the result of tracking the position and angle of a user seen from the first AP in a 4-user scenario. We can see that after an initial burst of sensing epochs, the EKF can track the user with high precision. We observe that the estimates deviate slightly from the true parameters until a new sensing epoch. After the initial sensing blocks, the EKF can track the user with very little error.

\section{Conclusion}

In this paper, we proposed an EKF-based predictive beamforming framework for cell-free massive MIMO that replaces conventional request-based channel estimation with persistent state tracking, enabling overhead-free downlink communication. The proposed variance-threshold-driven sensing management triggers sensing in approximately 1$\%$ of the epochs, yet sustains spectral efficiencies within a narrow margin of the perfect-CSI bound, demonstrating that continuous sensing is unnecessary once the EKF converges after a short initial burst. The distributed AP topology is instrumental to this robustness, unlike a co-located array, whose estimation quality degrades as the user moves away, the cell-free architecture provides uniform spectral efficiency. The MMSE precoder proves critical in translating imperfect state estimates into reliable performance, significantly outperforming MR and ZF alternatives by jointly suppressing multi-user interference and accounting for estimation uncertainty, even when users are closely spaced. Among waveform strategies, dedicated sensing waveforms yield the highest accuracy; however, allocating as few as 5$\%$ of subcarriers to coded radar pulses achieves comparable performance, offering a practical design point under spectral constraints. These findings establish that proactive, state-driven sensing management coupled with cell-free massive MIMO effectively decouples sensing and communication in time while maintaining high spectral efficiency, making the framework particularly attractive for high-mobility vehicular scenarios.


\appendix
\section{Appendix}
\subsection{The threshold for covariance of angle estimate} \label{appendix:The threshold for covariance of angle estimate}
The angle estimates from the EKF algorithm can be written as
\begin{equation}
\hat{\theta} \sim \mathcal{N}(\theta_0, \operatorname{var}(\tilde{\theta})),
\end{equation}
where $\theta_0$ is the mean value of the estimate and since $\hat{\theta} - \theta_0$ is normally distributed with zero mean and variance $\operatorname{var}(\tilde{\theta})$.
We wish to ensure that
$
\Pr\Bigl(|\hat{\theta} - \theta_0| > \theta_{\mathrm{HPBW}}\Bigr) < \epsilon$ for the half-power beamwidth $\theta_{\mathrm{HPBW}}$. We can define the standardized variable
$
\frac{\hat{\theta} - \theta_0}{\sqrt{\operatorname{var}(\tilde{\theta})}} \sim \mathcal{N}(0,1)
$ and write
\begin{equation}
\Pr\Bigl(|\hat{\theta} - \theta_0| > \theta_{\mathrm{HPBW}}\Bigr)
=2\left[1-\Phi\left(\frac{\theta_{\mathrm{HPBW}}}{\sqrt{\operatorname{var}(\tilde{\theta})}}\right)\right],
\end{equation}
where $\Phi(\cdot)$ denotes the cumulative distribution function of the standard normal distribution. Rearranging and solving for $\operatorname{var}(\tilde{\theta})$, the threshold on the covariance is then given by
\begin{equation}
\operatorname{var}(\tilde{\theta}) < \left(\frac{\theta_{\mathrm{HPBW}}}{\Phi^{-1}\Bigl(1-\frac{\epsilon}{2}\Bigr)}\right)^2=\gamma.
\end{equation}
The right side of the relation is the threshold $\gamma$.

\bibliographystyle{IEEEtran}
\bibliography{IEEEabrv,isac}

\begin{thebibliography}{10}
\providecommand{\url}[1]{#1}
\csname url@samestyle\endcsname
\providecommand{\newblock}{\relax}
\providecommand{\bibinfo}[2]{#2}
\providecommand{\BIBentrySTDinterwordspacing}{\spaceskip=0pt\relax}
\providecommand{\BIBentryALTinterwordstretchfactor}{4}
\providecommand{\BIBentryALTinterwordspacing}{\spaceskip=\fontdimen2\font plus
\BIBentryALTinterwordstretchfactor\fontdimen3\font minus \fontdimen4\font\relax}
\providecommand{\BIBforeignlanguage}[2]{{%
\expandafter\ifx\csname l@#1\endcsname\relax
\typeout{** WARNING: IEEEtran.bst: No hyphenation pattern has been}%
\typeout{** loaded for the language `#1'. Using the pattern for}%
\typeout{** the default language instead.}%
\else
\language=\csname l@#1\endcsname
\fi
#2}}
\providecommand{\BIBdecl}{\relax}
\BIBdecl

\bibitem{10217169}
H.~Hua, T.~X. Han, and J.~Xu, ``{MIMO} integrated sensing and communication: {CRB}-rate tradeoff,'' \emph{{IEEE} Trans. Wireless Commun.}, vol.~23, no.~4, pp. 2839--2854, 2024.

\bibitem{9737357}
F.~Liu, Y.~Cui, C.~Masouros, J.~Xu, T.~X. Han, Y.~C. Eldar, and S.~Buzzi, ``Integrated sensing and communications: Toward dual-functional wireless networks for {6G} and beyond,'' \emph{{IEEE} J. Sel. Areas Commun.}, vol.~40, no.~6, pp. 1728--1767, 2022.

\bibitem{9246715}
W.~Yuan, F.~Liu, C.~Masouros, J.~Yuan, D.~W.~K. Ng, and N.~González-Prelcic, ``Bayesian predictive beamforming for vehicular networks: A low-overhead joint radar-communication approach,'' \emph{{IEEE} Trans. Wireless Commun.}, vol.~20, no.~3, pp. 1442--1456, 2021.

\bibitem{9171304}
F.~Liu, W.~Yuan, C.~Masouros, and J.~Yuan, ``Radar-assisted predictive beamforming for vehicular links: Communication served by sensing,'' \emph{{IEEE} Trans. Wireless Commun.}, vol.~19, no.~11, pp. 7704--7719, 2020.

\bibitem{10214383}
Y.~Cui, Q.~Zhang, Z.~Feng, Q.~Wen, Z.~Wei, F.~Liu, and P.~Zhang, ``Seeing is not always believing: {ISAC}-assisted predictive beam tracking in multipath channels,'' \emph{{IEEE} Wireless Commun. Lett.}, vol.~13, no.~1, pp. 14--18, 2024.

\bibitem{khan2014localization}
R.~Khan, S.~U. Khan, S.~Khan, and M.~U.~A. Khan, ``Localization performance evaluation of extended {K}alman filter in wireless sensors network,'' \emph{Procedia Computer Science}, vol.~32, pp. 117--124, 2014.

\bibitem{du2022sensing}
Z.~Du, F.~Liu, and Z.~Zhang, ``Sensing-assisted beam tracking in {V2I} networks: Extended target case,'' in \emph{Proc. IEEE Int. Conf. Acoust., Speech, Signal Process. (ICASSP)}, 2022, pp. 8727--8731.

\bibitem{10278781}
Y.~Zhao, X.~Xu, Y.~Zeng, and F.~Liu, ``Sensing-assisted predictive beamforming with {NLoS} identification,'' in \emph{Proc. IEEE Int. Conf. Commun. (ICC)}, 2023, pp. 6455--6460.

\bibitem{10872824}
Y.~Zhao, X.~Xu, Y.~Zeng, F.~Liu, Y.~Huang, and Y.~L. Guan, ``Sensing-assisted predictive beamforming with multipath echo signals,'' \emph{IEEE Trans. Veh. Technol.}, vol.~74, no.~5, pp. 7539--7553, 2025.

\bibitem{9791349}
C.~Liu, W.~Yuan, S.~Li, X.~Liu, H.~Li, D.~W.~K. Ng, and Y.~Li, ``Learning-based predictive beamforming for integrated sensing and communication in vehicular networks,'' \emph{{IEEE} J. Sel. Areas Commun.}, vol.~40, no.~8, pp. 2317--2334, 2022.

\bibitem{9492131}
J.~Mu, Y.~Gong, F.~Zhang, Y.~Cui, F.~Zheng, and X.~Jing, ``Integrated sensing and communication-enabled predictive beamforming with deep learning in vehicular networks,'' \emph{{IEEE} Commun. Lett.}, vol.~25, no.~10, pp. 3301--3304, 2021.

\bibitem{akçalı2025predictivebeamformingdistributedmimo}
H.~T. Ak{\c c}al{\i}, {\"O}.~T. Demir, and T.~Girici, ``Predictive beamforming with distributed mimo and nlos identification,'' in \emph{Proc. Int. Conf. Smart Appl., Commun., Netw. (SmartNets)}, 2025, pp. 1--6.

\bibitem{5776640}
C.~Sturm and W.~Wiesbeck, ``Waveform design and signal processing aspects for fusion of wireless communications and radar sensing,'' \emph{Proc. {IEEE}}, vol.~99, no.~7, pp. 1236--1259, 2011.

\bibitem{9354629}
T.~Wild, V.~Braun, and H.~Viswanathan, ``Joint design of communication and sensing for beyond {5G} and {6G} systems,'' \emph{IEEE Access}, vol.~9, pp. 30\,845--30\,857, 2021.

\bibitem{7970102}
Y.~Liu, G.~Liao, J.~Xu, Z.~Yang, and Y.~Zhang, ``Adaptive {OFDM} integrated radar and communications waveform design based on information theory,'' \emph{{IEEE} Commun. Lett.}, vol.~21, no.~10, pp. 2174--2177, 2017.

\bibitem{9420261}
M.~F. Keskin, V.~Koivunen, and H.~Wymeersch, ``Limited feedforward waveform design for {OFDM} dual-functional radar-communications,'' \emph{{IEEE} Trans. Signal Process.}, vol.~69, pp. 2955--2970, 2021.

\bibitem{10548861}
S.~Mura, D.~Tagliaferri, M.~Mizmizi, U.~Spagnolini, and A.~Petropulu, ``Waveform design for {OFDM}-based {ISAC} systems under resource occupancy constraint,'' in \emph{Proc. IEEE Radar Conf. (RadarConf)}, 2024, pp. 1--6.

\bibitem{10742291}
U.~Demirhan and A.~Alkhateeb, ``Cell-free {ISAC} {MIMO} systems: Joint sensing and communication beamforming,'' \emph{{IEEE} Trans. Commun.}, vol.~73, no.~6, pp. 4454--4468, 2025.

\bibitem{yan2024communicate}
W.~Yan, O.~A. Topal, Z.~Behdad, {\"O}.~T. Demir, and C.~Cavdar, ``Communicate or sense? {AP} mode selection in mm{W}ave cell-free massive {MIMO}-{ISAC},'' in \emph{Proc. Asilomar Conf. Signals, Syst., Comput.}, 2024, pp. 889--893.

\bibitem{10901970}
M.~Elfiatoure, M.~Mohammadi, H.~Q. Ngo, H.~Shin, and M.~Matthaiou, ``Multiple-target detection in cell-free massive {MIMO}-assisted {ISAC},'' \emph{{IEEE} Trans. Wireless Commun.}, vol.~24, no.~5, pp. 4283--4298, 2025.

\bibitem{10681604}
Q.~Zou, Z.~Behdad, {\"O}.~Tu{\u g}fe~Demir, and C.~Cavdar, ``Distributed versus centralized sensing in cell-free massive {MIMO},'' \emph{{IEEE} Wireless Commun. Lett.}, vol.~13, no.~12, pp. 3345--3349, 2024.

\bibitem{du2025graph}
Y.~Du, S.~Xu, and J.~Chauhan, ``A graph-based hybrid beamforming framework for {MIMO} cell-free {ISAC} networks,'' \emph{arXiv preprint arXiv:2509.25385}, 2025.

\bibitem{zhang2025efficient}
J.~Zhang, S.~Xu, C.~Li, Y.~Huang, and L.~Yang, ``Efficient beam selection for {ISAC} in cell-free massive {MIMO} via digital twin-assisted deep reinforcement learning,'' \emph{arXiv preprint arXiv:2506.18560}, 2025.

\bibitem{11275289}
E.~B. Kama, M.~B. Salman, I.~Skog, and E.~Björnson, ``Sensing management for pilot-free predictive beamforming in cell-free massive mimo systems,'' in \emph{Proc. IEEE Int. Symp. Pers., Indoor, Mobile Radio Commun. (PIMRC)}, 2025, pp. 1--6.

\bibitem{ljung1987system}
L.~Ljung, \emph{System Identification: Theory for the User}, 2nd~ed.\hskip 1em plus 0.5em minus 0.4em\relax Upper Saddle River, NJ, USA: Prentice Hall, 1999.

\bibitem{dogandzic2001cramer}
A.~Dogandzic and A.~Nehorai, ``Cramer-{R}ao bounds for estimating range, velocity, and direction with an active array,'' \emph{{IEEE} Trans. Signal Process.}, vol.~49, no.~6, pp. 1122--1137, 2001.

\bibitem{1261132}
X.~Rong~Li and V.~Jilkov, ``Survey of maneuvering target tracking. part i. dynamic models,'' \emph{{IEEE} Trans. Aerosp. Electron. Syst.}, vol.~39, no.~4, pp. 1333--1364, 2003.

\bibitem{demir2021foundations}
{\"O}.~T. Demir, E.~Bj{\"o}rnson, and L.~Sanguinetti, ``Foundations of user-centric cell-free massive {MIMO},'' \emph{Foundations and Trends{\textregistered} in Signal Processing}, vol.~14, no. 3-4, pp. 162--472, 2021.

\end{thebibliography}

\end{document}